\documentclass[12pt]{article}
\usepackage{amsmath}
\usepackage{graphics, graphicx}

\def\refto#1{\ [#1]}
\def\half{{1/2}}

\def\prim{\prime}

\def\be{\begin{equation}}
\def\ee{\end{equation}}

\begin{document}

\title{Prediction in Quantum Cosmology}  
%\author{James B. Hartle,}
%\affiliation{Department of Physics\\
%University of California\\ Santa Barbara, CA 93106, USA}

\author{James B. Hartle \\
Department of Physics\\
University of California\\ Santa Barbara, CA 93106, USA}

\date{}

\maketitle

\centerline{\noindent\fbox{\parbox{5.5in}{
Lectures at the 1986 Carg\`ese summer school publshed in {\sl Gravitation in 
Astrophysics} ed.~by J.B.~Hartle  and B.~Carter,  Plenum Press,
New York (1987). In this version  some references that were `to be published' in the original have been supplied, a few typos corrected, but the text is only modestly edited. The author's views on  the quantum mechanics of cosmology have changed in important ways from those presented in Section 2. (See, e.g.  J.B.~Hartle, {\it Spacetime Quantum Mechanics and the Quantum Mechanics of Spacetime} in {\sl 
Gravitation and Quantizations}, Proceedings of the 1992 Les Houches
Summer School, ed.~by B.~Julia and J.~Zinn-Justin, Les Houches Summer
School Proceedings Vol.~LVII, North Holland, Amsterdam (1995).  The material in Sections 3 and 4 on the classical geometry limit and the approximation of quantum field theory in curved spacetime may still be of use.
}} }

\section{Introduction}
                  
As far as we know them, the fundamental laws of physics are quantum mechanical
in nature.  If these laws apply to the universe as a whole, then there must
be a description of the universe in quantum mechancal terms.  Even our
present cosmological observations require such a description in principle,
though in practice these observations are so limited and crude that the
approximation of classical physics is entirely adequate.  In the early
universe, however, the classical approximation is unlikely to be valid.
There, towards the big bang singularity, at curvatures characterized by the
Planck length, $(\hbar G/c^3)^{1\over 2}$, quantum fluctuations become important
and eventually dominant.

Were the aim of cosmology only to describe the present universe, expressing
that description in quantum mechanical terms might be an interesting intellectual
exercise but of no observational relevance.  Today, however, we have a more
ambitious aim:  to explain the presently observed universe by a simple and
compelling laws of its quantum state and dynamics .  It is natural to expect such  laws
to be quantum mechanical for several reasons:  The laws 
must describe the early universe where quantum gravitational fluctuations
are important.  In quantum fluctuations we can imagine a simple origin of
present complexity.  Finally, if all the other fundamental laws of physics
are quantum mechanical, it is only natural to expect the laws of initial
conditions and dynamics to be so also.  It is for the search for these 
that we need a quantum mechanical description of the universe -- a quantum
cosmology\refto{1,2-5}.  The nature of this description is the subject of
these lectures\refto{6}.

In the application of quantum mechanics to the universe as a whole, one
confronts the characteristic features of quantum theory in a striking and
unavoidable manner.  Some find these features uncomfortable, or unsatisfactory, 
or even absurd.  It is not the purpose of these lectures to examine whether
these attitudes represent a success or failure of intuition.  Rather, the
purpose is to sketch how the standard quantum theory, or a suitable
generalization of it, can be used to frame a law of initial conditions and
to extract from it predictions for cosmological observation.  We shall thus
assume quantum mechanics.

We shall also assume spacetime.  While it is uncertain whether Einstein's
 vision
of spacetime as a fundamental dynamical quantity is correct, it is perhaps
the most compelling viewpoint in which to frame a quantum theory of cosmology.
  Within this framework
of quantum spacetime our discussion will, in large part, be general and
not single out any particular dynamics or particular theory of initial
conditions.

The aim of these lectures is to show how the single system which is our
universe is described in a quantum theory of spacetime and to sketch how
a prescription for the quantum state of the universe can be used to make
verifiable predictions.  This is discussed in general in Sections 2 and
3.  However, as the observations which are accessible to us are describable
in classical terms, extracting predictions in the classical limit is a
particular problem of special importance.  This is considered in Section
4.  As an interesting by product of this discussion, we describe the connection
between a quantum theory of spacetime and the approximation of quantum field
theory in curved spacetime.
\def\3y{{^3{\cal G}}}
\def\4y{{^4{\cal G}}}
\def\cC{{\cal C}} 

\section{ Predictions of  the Wave Function of the Universe}
 
\subsection{The Wave Function of the Universe}
 
In quantum mechanics we describe the state of a system by giving
its wave function.  The wave function enables us to made
predictions about observations made on a spacelike surface; it
thus captures quantum mechanically the classical notion of the
``state of the system at a moment of time."  The arguments of
the wave function are the variables describing how the system's
history intersects the spacelike surface.  For example, for the
quantum mechanics of a particle, the histories are particle paths
$x(t)$.  We write for the wave function
$$
\psi=\psi(x,t).\eqno(2.1)
$$
The $t$ labels the hypersurface and the $x$ specifies the
intersection of the history with it.
 
In the quantum mechanics of a closed cosmologies with fixed (for
simplicity) spatial topology, say that of a 3-sphere $S^3$, the
histories are the 4-geometries, $^4{\cal G}$, on ${\bf R} \times S^3$.  The
appropriate notion of a 4-geometry fixed on a spacelike surface
is the 3-geometry, $\3y$, induced on that surface.  One can think
of this as specified by a 3-metric $h_{ij}({\bf x})$ on the fixed spatial
topology.  Thus for the quantum mechanics of a closed cosmology we
write{\refto 7}          
$$
\Psi=\Psi[\3y]=\Psi[h_{ij}({\bf x})].\eqno(2.2)
$$
Note that there is no additional ``time" label.  This is
because a generic 3-geometry will fit into a generic 4-geometry at
locally only one place if it fits at all.  The 3-geometry itself
carries the information about its location in spacetime.  For example, the
4-geometry of a closed Friedman universe is described by the metric
$$
ds^2 = -d\tau^2 + a^2(\tau)d\Omega^2_3,\eqno(2.3)
$$
where $d\Omega^2_3$ is the metric on the round 3-sphere.  A 3-geometry is
described by the radius of a 3-sphere.   This radius locates us locally
in the spacetime although in the large there are two values of $\tau$ for each
value of $a$.  This
labeling of the wave function correctly counts the degrees of
freedom.
Of the six components of $h_{ij}$, three are gauge. If one of
the remaining three is time, there are left two degrees of
freedom --- the correct number for the massless, spin-2 gravitational field.
 
The space of all three geometries is called superspace.  Each
``point" represents a different geometry on the fixed spatial
topology.  In the case of pure gravity that we have been
describing, the wave function is a complex function on
superspace.  With the inclusion of matter fields the wave
function depends on their configurations on the spacelike
surface as well, and we write typically
$$
\Psi=\Psi[h_{ij}({\bf x}),\Phi({\bf x})].\eqno(2.4)
$$
A law for initial conditions in quantum cosmology is a law which
prescribes this wave function.
 
\subsection{Cosmological Observations and Cosmological Predictions}
 
To make contact with observations we must specifiy the
observational consequences of the state of the universe
being described by this or another wave function.  This is
usually called an ``interpretation" of $\Psi$.  There is little
doubt that what I can say here will not address every issue
which can be raised on this fascinating topic and even less doubt
that it will not satisfy many who have thought about the
subject.  I would like, however, to offer some minimal elements
of an interpretation which I believe will enable an attribution
of $\Psi$ to the universe to be confronted by cosmological
observations.  These elements are an example of ``an Everett
interpretation" although the words and emphasis may be different
from other interpretations in this broad catagory{\refto 8}.

There are at least three problems to be addressed (1) the special nature
of cosmological predictions, (2) the quantum mechanics of single systems,
and (3) the problem of time.  We shall discuss them in order.

\subsection{The Nature of Cosmological Predictions}

 The favorite paradigm of prediction in
physics is evolution:  If we start the system in a certain state then a
time $t$ later we predict that it will be in a certain other state.  This
type of problem conforms to the characteristic form of predictive statements.
If ``this'' then ``that''-- a correlation between experimental conditions
and observations.  In placing conditions and observations in temporal
order, however, it is very uncharacteristic of predictions we can
make in the astrophysical and geological sciences.  In geology we might
predict as follows:  ``If we are in a certain type of strata, then we should
find a certain type of dinosaur bone.''  ``If we are in the middle of an
ocean floor, then we predict an upwelling trench'' and so on.  Here condition
and observation are at the same time.  This, prediction of correlations
{\it at the same time} is, I believe, characteristic of systems over which we
 have little experimental control.

Cosmology is much the same.  We can, of course, imagine a $10^9$ year
experiment ``Given the observations of the positions of the galaxies now, we
 predict
that if we wait $10^9$ years, we will see them in new positions $\cdot\cdot
\cdot\cdot$''  Such a prediction {\it is} a test of a theory of initial
 conditions
because the longer we wait the more initial data we see.  It is, however, 
not very practical and therefore
not very interesting.                

A more interesting type of prediction is a
$10^9 $ franc experiment:  Suppose you are allocated $10^9$ francs to
build new optical, radio, X-Ray, neutrino and gravitational wave telescopes, what
do you predict you will see?  A typical prediction might be the following:
``Given the locally measured values of the Hubble constant and the mass
density, we predict that at great distances we will see the same uniform mass
 density,
a certain galaxy-galaxy correlation function, 
a certain gravitational wave background, etc.''
Characteristically these predictions                                   
involve conditions and observations at a single moment
of time.
       
\subsection{Quantum Mechanics of Individual Systems}

The idea for dealing with the universe as a single system is to take quantum
mechanics seriously.  One assumes that there is one wave function $\Psi$
defined on a preferred configuation space which contains all the predictable
information about observations made on a spacelike surface.  {\it If $\Psi$ is
sufficiently peaked about some region in the configuration space, we predict
that we will observe the correlations between the observables which
characterize this region.  If $\Psi$ is small in some region, we predict
that observations of the correlations which characterize this region are
precluded.  Where $\Psi$ is neither small nor sufficiently peaked we don't
predict anything.}  That's it.

The natural reaction to such a proposal for interpretation is to ask ``Where is
probability?''  In response, two things can be said.
First, probabilities for single systems have no direct observational
meaning and the universe, by definition, is a single system.  Second,
as we shall show below,
this interpretation implies the usual probability interpretation of ordinary
quantum
mechanics when applied to ensembles of identically prepared systems.

In cosmology, therefore, we would examine any particular proposal for $\Psi$
to see which correlations are predicted -- those on which the wave function
is sharply peaked, and which are precluded -- those on which it is essentially
zero.  We would ask, for example:  ``Given the value of the Hubble constant and
the local mean density is the wave function sharply peaked about a form
of the galaxy-galaxy correlation function?''  If so we predict that correlation
of variables.  Note that characteristically we have conditions and observations
on a single spacelike surface.  This type of interpretation means that one's
ability to predict in quantum cosmology is very limited.  Given a value
of the Hubble constant and the local mean mass density, one can ask whether
the wave function
sharply peaked about the number of planets in the solar system, or the
architecture of this building, or the weights of the participants of this
conference.  I, for one, hope not.  One of the central problems in quantum
cosmology is therefore to find what correlations are predicted and how specific
must we be in conditions to get predictions for interesting observations.
(Problem 1).

Ordinary quantum mechanics can be formulated as a theory of individual 
systems.  Indeed, a moments reflection will show that this has to be so.
 Quantum mechanics formulated only in terms of probabilities would make
definite predictions only about infinite ensembles -- an idealization we
do not encounter in the real world.  Any ensemble can be regarded
as a single system composed of many identical parts.  Quantum mechanics
should be formulable as a theory of individual systems and the probability
interpretation derivable from the predictions this formulation makes about
single systems with many identical subsystems.  
In the late '60's a number
of workers independently showed how to do this{\refto 9}:

Consider a
single system and let it be described by a wave function $\psi$.  Possible
 observations
correspond to operators in the Hilbert space of states.  For the physical
interpretation of $\psi$ for an individual system assume only the following:
 {\it If $\psi$ is an eigenfunction of an observable $A$ then an observation
of $A$ will yield the eigenvalue.  For those observables
of which $\psi$ is not an eigenfunction there is no prediction for the outcome
of an observation.}  We can then derive the probability interpretation of
$\psi$ as follows:

Suppose the configuration space of the single system is $C$; the configuration
space for an ensemble of $N$ systems is $C^N$.  An ensemble of $N$ systems
each in the identical state $\psi (q)$ is described by the wave function
on $C^N$
$$
\Psi(q_1,\cdot\cdot\cdot,q_n) = \psi(q_1)\psi(q_2)\cdot\cdot\cdot\psi(q_n).
\eqno(2.5)
$$                                                                            
On the Hilbert space of wave functions on ${C}^N$ there is
an operator $\hat f_a$ corresponding to observing $q$ on the first system,
$q$ on the second, etc., and then computing the frequency that a given value
$a$ occurs.  For an infinitely large ensemble of identical systems, each
in a state $\psi$, is it a mathematical fact that the product wave function
(2.5) is an eigenfunction of this operator
$$\hat f_a \Psi = |\psi(a)|^2 \Psi.\eqno(2.6)$$
The predicted frequency is the square of the wave function of the single
system.
In this way we deduce the probability interpretation of quantum mechanics
from its predictions about individual systems.

To see how this works let us consider a definite example.  Consider ensembles
of spin -1/2 systems.  A single system has states $|S>, S=\uparrow$ or 
$\downarrow$, and the
Hilbert space of an ensemble of $N$ systems is spanned by the basis
$$
|S_1>|S_2>\cdot\cdot\cdot|S_N>.\eqno(2.7)
$$
In this basis we can define the operator corresponding to a measurement
of the relative frequency of say spin up, $\uparrow$
$$
f_{\uparrow}^N = \sum_{S_1\cdot\cdot\cdot S_N}|S_1>\cdot\cdot\cdot |S_N>
\Biggl ( \sum_{S_i}~\frac{\delta_{S_i\uparrow}}{N} \Biggr ) <S_N|
\cdot\cdot\cdot <S_1|.\eqno(2.8)
$$
Consider now the expectation value of $f_\uparrow^N$ in the state of an
ensemble of identically prepared systems each in state $|\psi>$
$$
|\psi^N> = |\psi>|\psi>\cdot\cdot\cdot |\psi>.\eqno(2.9)
$$
It is
\begin{align}
<\psi^N|f^N_\uparrow|\psi^N>&=\sum_{S_1\cdot\cdot\cdot S_N}
\Biggl ( \sum_{S_i} \frac{\delta_{\uparrow S_i}}{N} \Biggr ) |<S_1|\psi>|^2
\cdot\cdot\cdot|<S_N|\psi>|^2 \tag{2.11}\\
&=|<\uparrow|\psi>|^2 \sum_{S_2\cdot\cdot\cdot S_N}
|<S_2|\psi>|^2\cdot\cdot\cdot |<S_N|\psi>|^2
&= |<\uparrow|\psi>|^2. \nonumber 
\end{align}
Consider also the fluctuations about this mean value:
$$
<\psi^N|(f^N_\uparrow - |<\uparrow|\psi>|^2)^2|\psi^N>
 = <\psi^N|(f^N_\uparrow
)^2|\psi^N> - |<\uparrow|\psi>|^4.\eqno(2.12)
$$
The first term is
\begin{align}
\sum_{S_1\cdot\cdot\cdot S_N}& \Biggl ( \sum_{i=j}\frac{\delta_{\uparrow S_i}}
{N^2} + \sum_{i\not= j} \frac{\delta_{\uparrow S_i}S_{\uparrow S_j}}
{N^2}\Biggr )
|<S_1|\psi>|^2\cdot\cdot\cdot |<S_N|\psi>|^2\cr
& =\frac{|<\uparrow|\psi>|^2}{N} + |<\uparrow|\psi>|^4 \frac{N^2 - N}{N^2}.
\tag{2.13}
\end{align}
Thus
$$
<\psi^N|(f^N_{\uparrow} - |<\uparrow|\psi>|^2)^2|\psi^N> = \frac{1}{N}(|<\uparrow
|\psi>|^2 - |<\uparrow|\psi>|^4) \rightarrow 0~ {\rm as}~ N \rightarrow
\infty  \eqno{(2.14)}
$$
and we have indeed shown
$$
||f^N_{\uparrow}|\psi^N> - |<\uparrow|\psi>|^2|\psi^N>|| \rightarrow 0,
\eqno(2.15)
$$
which is (2.6).

The above derivation of the probability interpretation of ordinary quantum
mechanics can be cast into the language used to interpret the cosmological
wave function.  ``Superspace'' is the configuration space $C^N$ of the 
ensemble.  Equations (2.10) and (2.15) show that for large $N$ the wave
function of an ensemble of identical systems, (2.5), is increasingly sharply
peaked in the variable corresponding to a measurement of the frequency of
spin - $\uparrow$ in the ensemble.  The value about which it is peaked is
$|<\uparrow|\psi>|^2$.  Thus, we predict from the $\Psi$ of (2.5)
 that a 
measurement of the frequency should yield this value.

This correspondance in language, however, points up an incompleteness in
the interpretation of the cosmological wave function.  To give a precise
meaning to ``sufficiently peaked'' a measure is needed on configuration
space.  In ordinary quantum mechanics this is supplied by Hilbert space
as the above derivation shows.  However, there is no satisfactory Hilbert
space formulation of quantum gravity.  The reason is the problem of time.

\subsection{The Problem of Time}

Time plays a central and peculiar role in the formulation of Hamiltonian
quantum mechanics.  The scalar product specifying the Hilbert space of states
is defined at one instant of time.  States specify directly the probabilities
of observations carried out at one instant of time.  Time is the sole
observable not represented by an operator in Hilbert space but rather enters
the theory as a parameter describing evolution.  In the construction of
a quantum theory for a specific system, the identification of the time variable
is a central issue.

In non-relativistic classical physics time plays a special role which is
unambiguously transferred to quantum mechanics.  In special relativistic
quantum mechanics there is already an issue of the choice of time variable,
but there is also a resolution.  We can construct quantum mechanics using
as the peculiar time variable the time associated with a particular Lorentz
frame.  The issue is whether the quantum theory, so constructed, is consistent
with the equivalence of Lorentz frames.  It is.  There is a unitary relation
between the quantum theories constructed in different Lorentz frames and
physical amplitudes are therefore Lorentz invariant.

For the construction of quantum theories of spacetime the choice of time
becomes a fundamental difficulty.  A preferred foliation of spacetime by
spacelike surfaces is necessary to formulate canonical quantum mechanics.
 The classical theory certainly singles out no such foliation, and we have
no evidence that theories formulated on two different foliations are unitarily
equivalent.  There is thus a conflict between canonical quantum mechanics
and general covariance.

To resolve this conflict  we have, it seems to me, two choices
(1) Modify general relativity so at the quantum mechanical level
a preferred time is singled out, or
(2) Modify quantum mechanics so it does not need a
preferred time.

The first option has been much discussed.  In these lectures I would like
to offer a few thoughts about the second.

Feynman's sum over histories formulation of quantum mechanics is a natural
alternative starting point for constructing quantum theories of spacetime
in which the problem of time is neither as immediate nor as central as it is
in Hamiltonian quantum mechanics.  The basic ingredients of a sum over
histories formulation are these:

\begin{description}
\item{(1)}  {\it The Histories}:  A history ${\cal H}$ is the set of observables
which describe all possible experiments.  Examples are the particle paths
of ordinary quantum mechanics or the 4-geometries of spacetime physics.

\item{(2)}  {\it The Probability Amplitude for a History}.
The joint probability amplitude for making all the
observations which make up a history is
$$\Phi({\cal H}) = \exp [iS({\cal H})],\eqno(2.16)$$
where $S({\cal H})$ is the classical action for the history.  For a particle this
is
$$S[X(\tau)] = \int^{\tau^{\prime\prime}}_{\tau^{\prime}} d\tau
\Bigl [\frac12 m\dot X^2 - V(X)\Bigr ],\eqno(2.17)$$
or for spacetime with the dynamics of general relativity it is
$$S[^4{\cal G}] = \frac{1}{\ell^2}\int_M d^4x R\sqrt{-g}    
  -\frac{2}{\ell^2} \int_{\partial M} d^3x h^{1\over 2}K.\eqno(2.18)$$

\item{(3)}  {\it Conditional Probability Amplitudes}.
         In particular experiments the observables can be divided
into three classes
                 (i)  The conditions ${\cal C}$ -- 
                      those observables fixed 
         by the experimental arrangement.     
                 (ii)                The observations ${\cal O}$ -- the results
of the experiments.
                 (iii)  The unobserved ${\cal U}$ -- those observables neither
conditioned nor observed.
The conditional amplitude for ${\cal O}$ given ${\cal C}$ is (the principle
of superposition)
$$\Phi ({\cal O} | {\cal C}) = \sum_{\cal U} \Phi ({\cal U}).\eqno(2.19)$$
A measure, which is just as important as the action, is needed to define
such sums.

\item{(4)}  {\it Probability}.
The relative probability that ${\cal O}$ occurs in a set of observations
given the conditions ${\cal C}$ is $|\Phi ({\cal O} | {\cal C}) |^2$.  From
this, the probabilities of one outcome of an exhaustive and exclusive set
of observations can be computed by appropriate normalization according to
the usual rules for probability.
\end{description}

Some further restrictions and caveats must be given, but this is the basic
framework.  The ideas will perhaps become clearer with an example illustrated
in Figure 1.
We consider a non-relativistic particle moving in one dimension.  Suppose
the particle starts at $X_1$ at time $\tau_1$.  It then passes through a
slit of width $\Delta_2$ at time $\tau_2$.  At $\tau_3$ there is a coherent
detector which registers whether the particle is in the interval $\Delta_3$
disturbing the particle as little as possible.  Finally at $\tau_4$ the
particle's position is detected.  The conditions ${\cal C}$ for this experiment
might be those imposed at times $\tau_1$ and $\tau_2$.  A complete and
exclusive set of observations would then be, $X_4$, the value of the position
at $\tau_4$ {\it and} whether the detector at $\tau_3$ registered (``clicked'')
or did not.  Given the experimental arrangement one of these possibilities
must happen and no more than one can happen.

\begin{figure}[t!] 
\centerline{\includegraphics[height=3in]{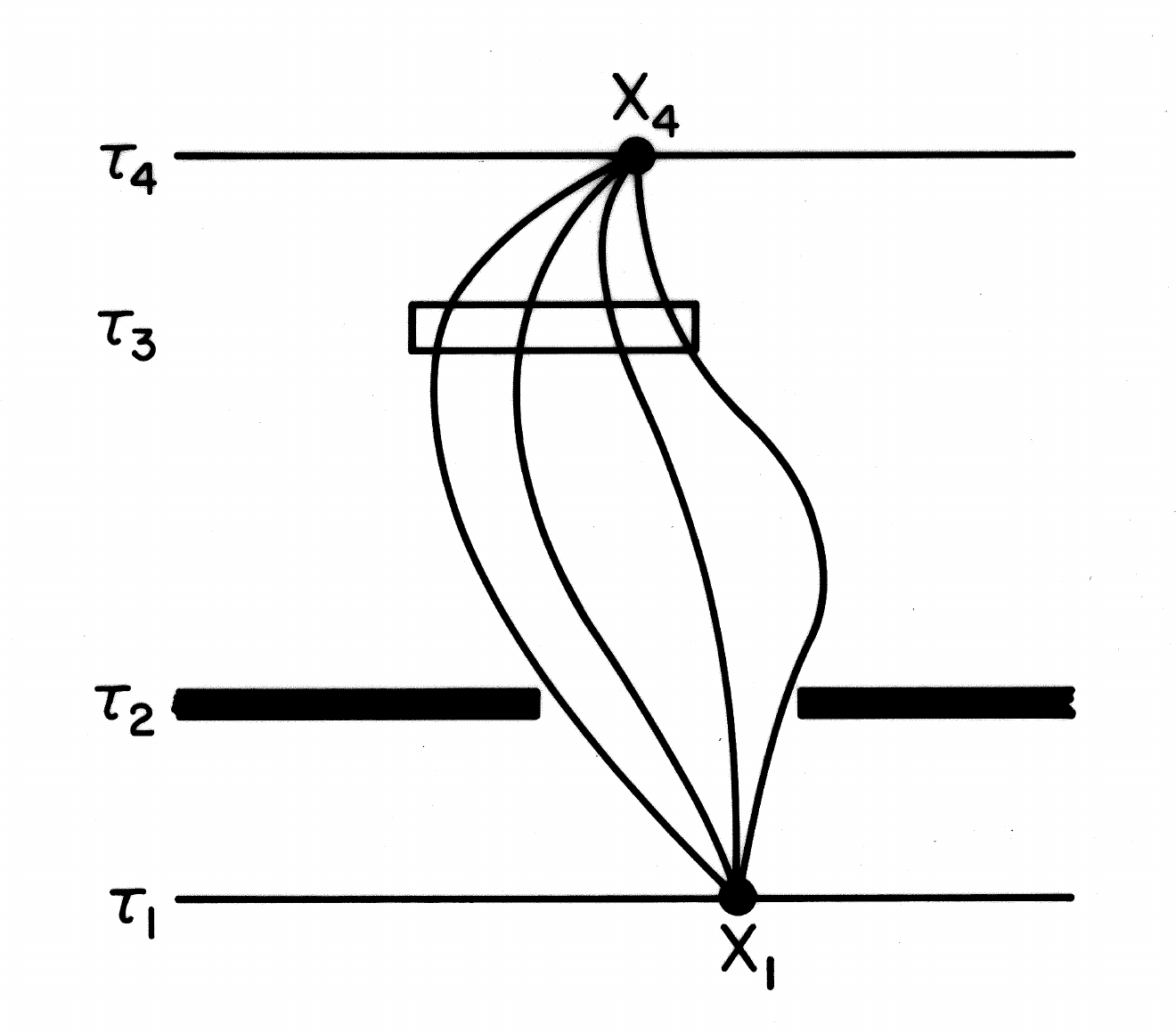}}
\caption{An example illustrating the construction of conditional
probability amplitudes.  A particle is localized at $\tau_1$, passes through
a slit at $\tau_2$, registers or does not register in a detector volume at
$\tau_3$, and its position is determined at $\tau_4$.  Given that the particle
started at $X_1$ and passed through the slit, the conditional probability
amplitude for it to register in the detector and be at $X_4$ is a sum over
all paths which start at $X_1$ pass through the slit and detector volume
and end at $X_4$.} 
\end{figure}                                                        

Let us compute the probability for the detector to click and the particle
to be found at $X_4$ in a range $\Delta_4$.  The conditional probability
amplitude $\Phi (X_4, {\rm click} |{\cal C})$ is the sum over all paths
which start at $X_1$ at $\tau_1$ pass through the slit at $\tau_2$, cross
the detector volume at $\tau_3$ and end at $\tau_4$ at $X_4$. (Figure 1).
 The conditional amplitude $\Phi (X_4,~ {\rm noclick}~|{\cal C})$
would be a similar sum over paths which cross outside the detector
volume at $\tau_3$.  The probability is the square of
 $\Phi (X_4,~ {\rm 
click}~|{\cal C})$
normalized over the set of
complete and exclusive possibilities.  That is
\begin{align}
P(X_4 & {\rm in} \Delta_4, {\rm click}|{\cal C}) = \Delta_4|\Phi(X_4,
 {\rm click}|{\cal C})|^2 \nonumber \\
&\times \Biggl [ \int_R dX_4 |\Phi(X_4, {\rm click}|{\cal C})|^2 +
\int_R dX_4|\Phi (X_4,{\rm noclick}|{\cal C})|^2\Biggr ]^{-1}. \tag{2.20}
\end{align}

The same result is predicted by ordinary quantum mechanics\refto{10}.
There one would say that the state at time $\tau_1$ was $|X_1\tau_1>$,
 the state
with the particle localized at $X_1$. 
After the localization at $\tau_2$ the ``wave packet is reduced'' and the
state is
$$|\psi_2\rangle \equiv N^{-2}_2 \int_{\Delta_2} | x_2\tau_2\rangle \langle
x_2\tau_1 | X_1 \tau_1\rangle,\eqno(2.21)$$
with $N_2$ determined so the state has unit norm.
           At $\tau_3$ the ``wave packet is again reduced.''  The
 probability that
the particle is inside the detector at $\tau_3$ is                 
$$P({\rm click}) = \int_{\Delta_3}dx_3 | \langle x_3\tau_3 | \psi_2\rangle
 |^2.
\eqno(2.22)$$  
The state after detection is the normalized projection of $|\psi_2>$ 
on the interval $\Delta_3$
$$|\psi_3\rangle = N^{-2}_3 \int_{\Delta_3} dX_3 |X_3\tau_3 \rangle \langle
X_3\tau_3 | \psi_2 \rangle.\eqno(2.23)$$
          At $\tau_4$ the probability that the particle is at $X_4$ and the
detector has clicked is the product 
$$\Delta_4 |\langle X_4\tau_4 | \psi_3 \rangle |^2 P({\rm click}).\eqno(2.25)$$
This is the same as (2.20) as an explicit calculation will show.

The contrast between the usual discussion of this experiment and that in
the sum over histories formulation shows that the sum over histories formalism
handles
                                  observations at different times
democratically and efficiently,
so that it is well adapted to deal with conditions and observations which
are not in temporal order.
In particular, the sum over histories formulation can deal efficiently and
naturally with observations which lie on a general surface, $\tau = \tau(X)$,
and with conditions which also lie on the surface.  It is thus especially
useful in cosmology.

We can now ask whether we can recover the Hilbert space formulation of quantum
mechanics from its sum over histories version.  This is easy to do on the
 surfaces of the preferred        
 non-relativistic time.  A conditional amplitude $\Phi ({\cal O}|{\cal C}) =
\sum_{paths}  e^{iS}$ for which the {\it conditions and observations are
temporally ordered}
can be factored into a sum over paths before an intermediate surface $\tau$ and a sum
after $\tau$ (Figure 2a)
$$\Phi ({\cal O} | {\cal C}) = \int dX \psi^*_{\cal O} (X\tau) \psi_{\cal C}
(X\tau),
\eqno(2.26)$$
where $\psi_{\cal C}(X\tau) = \sum_{paths~in~M_-}        e^{iS}$ 
for paths which meet the conditions ${\cal C}$ and end at $X$.  There is a similar
expression for $\psi^*_{\cal O}$.  This sum defines the wave function 
from the sum over paths.  Further, it has its usual probability interpretation
because the positions at $\tau$ are a set of exhaustive and exclusive
observations.

\begin{figure}[t!]
\centerline{\includegraphics[width=5in]{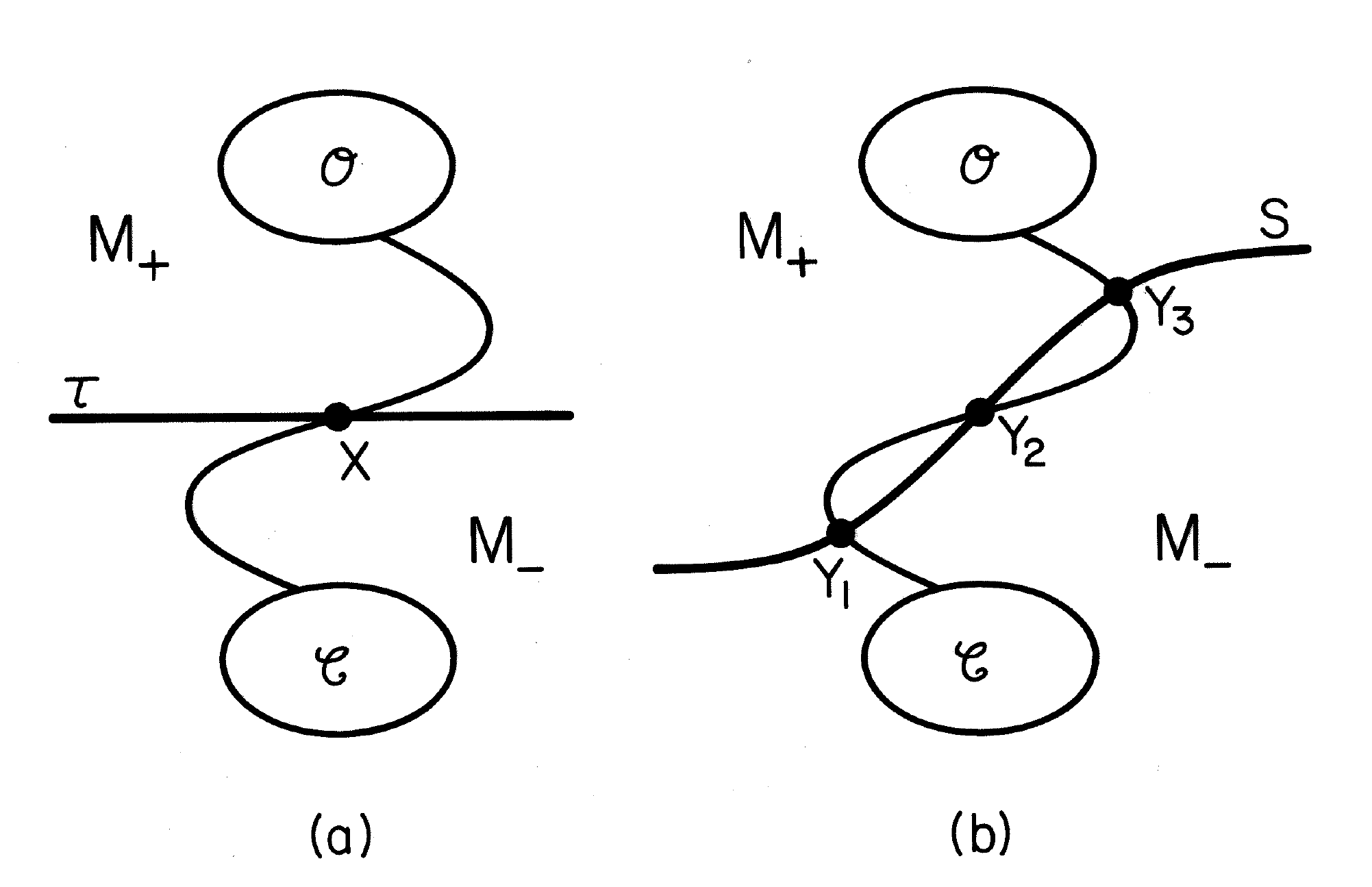}}
\caption{ (a)  When conditions, ${\cal C}$, and observations, 
${\cal O}$, are in temporal order, a path contributing to the conditional
amplitude $\Phi ({\cal O} | {\cal C})$ intersects an intermediate surface
of constant preferred non-relativistic time $\tau$ at one and only one
position $X$.  The sum defining the amplitude may thus be factored in to
a sum over paths prior to $\tau$, a sum over paths after $\tau$ and a sum
over $X$.  (b)  By contrast, a path contributing to the conditional amplitude
$\Phi ({\cal O} | {\cal C})$ may intersect a general surface $S$ many times. Indeed, the expected number of intersections is infinite.  }
 \end{figure}           

How would this construction go on a more general surface?  The crucial
difference is that now the paths can cross and recross the surface many
times.  (Figure 2b).  Formally, since each path is divided into parts by the
 surface,
one could write down

$$         \Phi ({\cal O} | {\cal C}) = \sum_{{{number~of}\choose {crossings,~
n}}} \int dY_1 \cdot\cdot\cdot dY_n
\psi^*_{\cal O} (Y_1\cdot\cdot\cdot Y_n, S) \psi_{\cal C} (Y_1\cdot\cdot\cdot
Y_n, S),\eqno(2.27)$$
where
$$\psi_{\cal C} = \sum_{paths~in~M_-}        e^{iS}.\eqno(2.28)$$
Some attention is now needed to define what one means by sums over regions
like $M_-$.  This can be done by introducing a spacetime lattice and going
to Euclidean time.
Then there is a close connection between sums over histories and the continuum
limits of stochastic processes.  In particular, for a free particle the sums
over histories can be defined as the continuum limit of a random walk.

When one calculates the sum for a fixed number of crossings in this manner,
one finds that the amplitude vanishes!  A composition law of the       form
 (2.27) exists on the lattice but does not have a continuum limit.  We do not
 recover
a Hilbert space formulation of quantum mechanics on a general surface 
$\tau = \tau (X).$
The reason is that the expected number of crossings is infinite and the
amplitude for any finite number of crossings is zero.  Only due to the peculiar
fact that the paths move forward in the preferred non-relativistic time
can we recover a Hilbert space formulation of the theory on surfaces of
that time because we know the paths cross them once and only once.

The absence of a Hilbert space for a general surface is not an obstacle to
the computation of probabilities for observations and conditions which lie
on the surface.  The sum over histories formulation allows this and our
specific example illustrates it.  However, typically, greater care is needed
to pose questions with sensible answers (unlike the amplitude of (2.28)).

Now imagine for a moment that we had been brought up on sum over histories
quantum mechanics.  What would our attitude be if we encountered a theory
in which there was {\it no} surface on which we could recover a Hamiltonian,
Hilbert space formulation of the theory?  Would we insist on this as a
necessary requirement for a successful quantum theory?  I would like to
suggest that the answer should be no.

I would now like to describe two examples of theories in which it seems
is not possible to recover a Hilbert space formulation, at least not by the
 methods we
have been describing.  The first is non-relativistic quantum mechanics with
real clocks.  The second is general relativity.

We have no direct perception of the time of non-relativistic mechanics.
 Psychological time is certainly poorly correlated with this variable. 
What we do observe are the positions of indicators in mechanical systems
carefully arranged so these indicators are correlated with the time of the
Schr\"odinger equation.  Such systems
are called clocks.  The simplest example is a free particle moving
with definite velocity in one dimension.  Its position $T$, appropriately
calibrated, gives the time.

In quantum mechanics an ideal clock would be one whose position $T$ remained
always correlated with the Schr$\ddot {\rm o}$dinger equation time $\tau$. 
 A possible
solution to the Schr$\ddot {\rm o}$dinger equation is then   
$$\psi (T,\tau) = \delta (T-\tau).\eqno(2.29)$$ 
The corresponding Hamiltonian would be linear in the momenta
$$h_C = - i      \frac{\partial}{\partial T} = P_T.\eqno(2.30)$$
Such clocks do not exist.  The energy of this Hamiltonian would be unbounded
below.

Real clocks, such as our free particle, can exist.  Their energy is positive
because their Hamiltonians are quadratic in the momenta
$$h_C = \frac{P_T^2}{2M}.\eqno(2.31)$$ 
However, such clocks are inevitably imperfect.  In the case of a particle
even if there is initially a sharp correlation between $T$ and $\tau$, eventually
the wave packet will spread and the clock will lose accuracy.  The spreading
can be reduced arbitrarily by making the particle sufficiently massive.
 Such general limitations on the masses of clocks have been discussed by
Salecker and Wigner\refto{11}.  These limitations are not important in non-relativistic
quantum mechanics because one can imagine arbitrarily massive clocks which
do not disturb the system.  In the gravitational physics of closed systems,
however, these limitations become fundamental.

Suppose that we try and construct a non-relativistic quantum mechanics in
which the indicators of real clocks are involved directly.  This is a problem
which has been discussed by DeWitt\refto{12}, Peres\refto{13}, Page and
 Wooters\refto{14} and no doubt
by many others.  From the sum over histories point of view one might proceed
as follows:
The histories are the world lines $X(\tau), T(\tau)$; moving forward in
$\tau$ but both forward and backward in $T$.  The observables and conditions
will involve $T$ and $X$ and therefore ${\it sums}$ over the unobserved
parameter $\tau$.
An interesting set of correlations are the values of $X$ for a given $T$.
                                        If a further part of the conditions
is to put the $T$ - component of the
system into ``a good clock state'' we recover an approximation to the
predictions of ordinary quantum mechanics.  By making the clocks massive
we can make this approximation as accurate as desired.  For no
finite mass, however, do we recover a Hilbert space formulation of the theory
on surfaces of constant $T$.  Formulated in terms of $T$ and $X$ there are
no preferred surfaces, and the paths cross and recross a given $T$     
an arbitrarily large number of times.

My last example is general relativity and, in particular, the quantum 
mechanics of closed cosmologies.  The histories, as will be described in
more detail in the next section, are cosmological 4-geometries.  There is
no natural time parameter for a cosmological history.  That is, there is
no parameter constructed from the metric and matter fields which, for a general
history, defines a foliating family of spacelike hypersurfaces such that
the parameter takes a distinct value on each surface.  Put differently,
for a candidate parameter such as $\sqrt h$ or $K$, one can find histories
that move forward and backward through a given value an arbitrarily large
number of times.  There is no preferred time.  It seems unlikely, therefore,
that one will recover a Hilbert space of states from sum over histories
quantum mechanics by the analog of straightforward construction used for
a particle earlier in this section.  I would like to suggest, however, that,
even in the absence of a Hilbert space formulation, one can formulate a
predictive quantum mechanics of the single system which is our universe
using the sum over histories formulation I have described.  Correlations
involving both conditions and observations on a spacelike surface can be
investigated using the measure supplied by the sum over histories.  Predictions
of correlations verifiable for the single system can be made if the wave
function is sufficiently peaked or sufficiently small.  This formulation
is but a slight modification of Hamiltonian quantum mechanics which coincides
with that formulation when a Hamiltonian theory is available.  In the following
sections we shall discuss some first steps towards the implementation of
this program.
  
\section{ Laws for Initial Conditions}

\subsection{The Sum Over Histories Formulation of Quantum 
Cosmology }

To apply the sum over histories formulation of quantum mechanics to cosmology
three things are needed:  the histories, the action, and the measure.  The
histories are cosmological spacetimes with matter fields.  For simplicity
we shall take the spacetimes to be spatially closed, Lorentzian 4-geometries
whose topology is of the form ${\bf R}  \times M^3$ with $M^3$ a compact
 3-manifold
(typically the 3-sphere).  There is no compelling reason for this restriction
on topology, and indeed it is interesting to investigate other possibilities
\refto{15}, but this assumption will simplify our discussion without limiting
the central ideas.

There is, today, no choice for the action of spacetime coupled to matter
which yields a satisfactory quantum field theory judged by familiar local
standards.  Whatever the correct theory, we expect that its low energy limit 
 will
be Einstein gravity coupled to matter.  The action for Einstein gravity
on a spacetime region $M$ is
$$\ell^2 S_E[g] =  \int_M d^4 x (-g)^{1\over 2}R - 2 \int_{\partial M} d^3
x (h)^{1\over 2} K.\eqno(3.1)$$
Here, $R$ is the scalar curvature, $\ell = (16 \pi G)^{1\over 2}$ is the
Planck length, $\partial M$ is the boundary of $M$, $h_{ij}$ is the metric
induced on the boundary by $g_{\alpha\beta}$ and $K$ is the trace of the
extrinsic curvature of the boundary.  The action for a free scalar field,
$\Phi$, with mass $M$ is a simple representative of the many possible matter
field actions,
\begin{align}
S_\Phi[g,\Phi] &= - \frac12 \int d^4x (-g)^{1\over 2} \Bigl
[ (\nabla\Phi)^2 + \xi R \Phi^2 + M^2 \Phi^2 \Bigr ]\cr
&+ \xi \int_{\partial M} d^3 x (h)^{1\over 2} K \Phi^2. \tag{3.2}
\end{align}
Where concrete illustrations of the action are needed, we shall use (3.1)
and (3.2).

Specification of the weights with which to carry out the sum over histories
is just as important for quantum mechanics as the specification of the action.
 The sums over paths defining the quantum mechanics of a particle may be
given a concrete meaning as the limit of sums over increasingly refined
piecewise linear approximations to those paths.  Weights can be assigned
to each piecewise linear path defining concretely a ``measure'' on the space
of paths.  Sums over geometries may be given concrete meaning as the limit
of sums over piecewise flat approximations to them using the methods of
the Regge Calculus\refto {16} and in this way a ``measure'' on geometries can be
defined\refto{17}.

Conditional probability amplitudes are formed from $\exp (iS)$ by summing
over geometries and field configurations.  We have argued that representative
predictions in cosmology involve observations made locally, ``at one moment
of time,'' with conditions specified, in part, at the same moment of time.
 Such predictions are extractable from the amplitude for observations on
a spacelike surface, that is, from the wave function.  The wave function
for a spacelike surface is determined by a sum over histories
restricted by conditions ``in the past'' of the spacelike
surface.  Specifically we write
$$\Psi_C[h_{ij}({\bf x}), \Phi ({\bf x}) ] = \int_C \delta g \delta \Phi \exp
\Bigl (iS[g, \Phi ] \Bigr ),\eqno(3.3)$$
where the sum is over cosmological 4-geometries and field configurations
which match the arguments of the wave function and satisfy the conditions
$C$.

The wave function on a spacelike surface is not the only conditional
probability amplitude which could be computed.  It is, however, the one
from which we expect to deduce most interesting cosmological predictions.
 We do this not by calculating probabilities, for the universe is a single
system.  Rather, as described in Section 2, we search the configuration
space for regions where the wave function is sharply peaked.  These
correlations are the predictions of quantum cosmology.  Given a set of
conditions $C$, it is a largely open, but important question, what predictions
one can expect.  As we shall describe in Section 4, this problem is greatly
simplified if there is a region of configuration space in which the wave
function can be approximated semiclassically.

A law for initial conditions in quantum cosmology is a law for the conditions
$C$.  That is, a law for initial conditions is a specification of the class
of geometries and field configuration which are summed over in (3.3) to
yield the wave function of the universe.

\subsection{Constraints}

We are not free to specify any wave function as a theory of initial conditions.
It must be representable as a sum over histories of the form (3.3) reflecting
the underlying gravitational dynamics.  In particular, it must satisfy certain
{\it constraints} which are consequences of this dynamics.  We shall now
briefly review these using the example of pure Einstein gravity\refto{7}.

There are four constraints in general relativity.  Three of them arise from
the requirement that the wave function, $\Psi [h_{ij}]$, depend only on three
geometry and not on the choice of coordinates used to describe that geometry.
$\Psi$ must thus be the same on two three
metrics which are connected by a diffeomorphism.  Infinitesimal diffeomorphisms
are generated by a vector $\xi^k$ according to
$$h_{ij} \rightarrow h_{ij} + D_{(i}\xi _{j)}.\eqno(3.4)$$
Thus, for infinitesimal $\xi^k$
$$\Psi [h_{ij} + D_{(i}\xi _{j)}] = \Psi [h_{ij}],\eqno(3.5)$$   
or equivalently
$$\int_{M^3} d^3 x D_{(i}\xi _{j)} \frac{\delta\Psi}{\delta h_{ij}({\bf x})} =
0.\eqno(3.6)$$
Integrating by parts on the compact manifold $M^3$, and recalling that $\xi^k$
is arbitrary, one arrives at the three constraints
$$D_i \Bigl (\frac{\delta \Psi}{\delta h_{ij}(\bf x)} \Bigr ) = 0.\eqno(3.7)$$
These are called the ``momentum'' constraints.

The fourth constraint of general relativity arises because general relativity
is an example of a parametrized theory in which time occurs as one of the
dynamical variables.  To illustrate the idea we begin with a simple model
\refto{7}.

Consider a non-relativistic particle whose dynamics is described by the
action
$$S[X(T)] = \int dT~\ell(dX/dT, X).\eqno(3.8)$$
Express both $X$ and $T$ as functions of a parameter $\tau$ and thereby
introduce the time $T$ as a dynamical variable in the action
$$S[X(\tau),T(\tau)] = \int d\tau \dot T \ell (\dot X/\dot T, X),\eqno(3.9)$$
where a dot denotes a $\tau$-derivative.  This action is invariant under
reparametrizations of the label time
$$\tau = f(\tau^{\prim}),~X^{\prim}(\tau^{\prim}) = X \bigl (f(\tau^{\prim})
 \bigr ),
~T^{\prim}(\tau^{\prim}) =
T \bigl (f(\tau^{\prim})\bigr ).\eqno(3.10)$$
If we calculate the Hamiltonian associated with the Lagrangian in (3.9),
we find first that
$$H = \dot T (p_T + h) ,\eqno(3.11)$$
where $p_T$ is the momentum conjugate to $T$ and $h$ is the Hamiltonian
associated with the Lagrangian $\ell$.  Second, we find that, identically,
$$H = 0.\eqno(3.12)$$
The vanishing of the Hamiltonian is a characteristic feature of theories
which are invariant under reparametrizations of the time.

In the quantum mechanics of this model, we construct the wave function $\psi_C(X,T)$
for a particular moment of time as a sum of $\exp (iS)$ over an appropriate
class of paths, $X(T)$.  We can carry out this sum in parametrized form
-- integrating over histories $X(\tau), T(\tau)$ and using the action (3.9).
However, histories which differ only by a reparametrization of $\tau$[eq.
(3.10)] correspond to the same path.  To count these only once in the sum
over histories we can ``fix'' the parametrization by requiring a particular
relation between $\tau$ and $T$
$$\tau = F(T),\eqno(3.13)$$
for arbitrary increasing $F(T)$ and write the sum over histories as
$$\psi_C(X,T) = \int_C \delta X \delta T \Biggl | \frac{dF}{dT} \Biggr | 
\delta \bigl (\tau - F(T) \bigr ) \exp \Bigl (iS[X,T] \Bigr ).\eqno(3.14)$$
The functional $\delta$ - function is the analog of the ``gauge fixing $\delta$
- function'' for gauge theories and $|dF/dT|$ is the analog of the
``Faddeev-Popov determinant.''

The familiar sum over histories for the quantum mechanics of a non-relativistic
particle is recovered from (3.14) by doing the integral over $T$ (most easily
by choosing $F = T$).  From this, and therefore from (3.14), the Schr\"odinger
equation follows.  Writing it in the form
$$\Biggl (-i\frac{\partial}{\partial T} + h \Biggr) \psi_C (X, T)
 = 0,\eqno(3.15)$$
we see that it is the operator form of the constraint $H = 0$.  Thus the
classical constraints arising from invariance under reparametrization
of the time are enforced as operator relations in quantum mechanics.  One
sees that the vanishing of the Hamiltonian for a parametrized theory does
not mean the absence of dynamics, it {\it is} the dynamical
relation.

General relativity is invariant under the group of diffeomorphisms in four
dimensions.  There are correspondingly four constraints.  They can be written
in ``$3 + 1$ form'' by choosing a family of spacelike surfaces 
and using as basic variables their intrinsic metric, $h_{ij}$, and extrinsic
curvature, $K_{ij}$.

Three of the four constraints express the invariance under diffeomorphisms
in the 3-surface.  These are the constraints we have already discussed.
 The fourth constraint expresses the invariance of the theory under choice
of the choice of spacelike surfaces, that is, under reparametrizations of
the time.  As in the simple model, the constraint is that the total
Hamiltonian (density) vanish.  For this reason it is called the 
{\it Hamiltonian constraint}.  Classically its form is
$$H = \ell^2 G_{ijk\ell} \pi^{ij} \pi^{k\ell} + \ell^{-2} h^{1\over 2}
\Bigl ( - ^3 R + 2 \Lambda \Bigr ) =0, \eqno (3.16)$$ 
where $^3 R$ is the scalar curvature of the 3-surface, $\pi^{ij}$ are
the momenta conjugate to $h_{ij}$
$$\ell^2\pi_{ij} = h^\half (K_{ij} - h_{ij} K)\eqno(3.17)$$
 and $G_{ijk\ell}$ is the ``supermetric''
$$G_{ijk\ell} = \frac12 h^{-{1\over 2}} \Bigl (h_{ik}h_{j\ell} + h_{i\ell}
h_{jk} - h_{ij} h_{k\ell}\Bigr ).\eqno(3.18)$$

Quantum mechanically eq (3.17) becomes an operator constraint on the wave
function called the Wheeler-DeWitt equation.  It takes the form
$$\Biggl [ - \ell^2 \nabla^2_x + \ell^{-2} h^{1\over 2} \Bigl ( - ^3 R +
2\Lambda \Bigr ) \Biggr ] \Psi \bigl [h_{ij}\bigr ] = 0,\eqno(3.19)$$ 
where
$$\nabla^2_x = G_{ijk\ell}~ \frac{\delta^2}{\delta h_{ij} ({\bf x}) \delta
h_{k\ell} ({\bf x})} + {{\rm linear~derivative~terms}\choose
 {\rm depending~on~factor~ordering}}.\eqno(3.20)$$
The Wheeler-DeWitt equation follows formally from the sum over histories
for quantum cosmology in much the same way that the Schr\"odinger equation
follows from the sum over parametrized paths (3.14) (Problem $2$).
It may be thought of as a functional differential equation which expresses
the dynamics of quantum cosmology in much the same way that the Schr\"odinger
equation expresses the dynamics of particle quantum mechanics.

\subsection{A Proposal for a Wave Function of the Universe}

There have been a number of proposals for a quantum state of the 
universe\refto{1-5}.  Perhaps the most developed of these is the proposal of
 Stephen
Hawking and his coworkers that the wave function of the universe is determined
by a sum over compact Euclidean 4 - geometries.  Detailed expositions of
this idea can be found elsewhere\refto{18, 6}.
Here we shall just state the proposal so that there is at least
one concrete idea with which to illustrate the subsequent discussion.

Euclidean sums over histories as well as Lorentzian ones may be used to
construct solutions to constraints.  Consider, for example, the sum over
particle paths
$$\psi_0 (X_0) = \int_{C_0} \delta X \exp \left (- I \left [ X(T) \right ]
\right),\eqno(3.21)$$
where $I$ is the Euclidean action for a non-relativistic particle in a
potential $V (X)$ 
$$I\bigl [ X(T) \bigr] = \int d T \Biggl [\frac12 M \Biggl({\frac{d X}{d
T}}\Biggr )^2 + V(X) \Biggr ],\eqno(3.22)$$
and the sum is over all paths which start at $X_0$ at Euclidean time $T
= 0$
and proceed to a configuration of minimum action at large negative times.
 The wave function $\psi_0 (X_0)$ so defined satisfies the constraint (3.15).
It is, in fact, the ground state wave function.

Euclidean sums over 4 - geometries give solutions to the operator constraints
of gravitational theories.  Consider a cosmological manifold of the form
${\bf R}^+ \times M^3$, where ${\bf R}^+$ is half the real line.  The manifold
 thus has an $M^3$
boundary.  A sum over Euclidean 4-geometries and field configurations of
the form
$$\Psi \Bigr [h_{ij} ({\bf x}), \Phi ({\bf x}) \Bigr ] = \int_C \delta g \delta
\Phi \exp \Bigl ( - I [g, \Phi ] \Bigr ),\eqno(3.23)$$
where $I$ is the Euclidean action for Einstein gravity coupled to matter,
will formally satisfy the constraints (3.7) and (3.9) provided the metric
and matter field induced on the boundary by each contributor to the sum
match those in the argument of the wave function\refto{28}.  (Problem 2).
 A particular wave function                                                          
is singled out by summing over {\it compact 4 - geometries with no other
boundary} and over {\it matter field configurations which are regular on these
geometries.}  The proposal of Hawking and his coworkers is that this
{\it is} the wave function of our universe.  

The Euclidean action for Einstein gravity
$$\ell^2 I[g] = - \int_M d^4 x g^\half R - 2 \int_{\partial M} d^3 x h^\half
K\eqno(3.24)$$
is not positive definite.  Thus, for general relativity and other theories
with this property the contour of integration in (3.23) cannot be over purely
real metrics -- the integral would diverge.  The contour of integration
must be taken in complex directions\refto{19}.  From the Hamiltonian
perspective one is free to make this distortion as long as the correct sum
over the true physical degrees of freedom is preserved.  This seems to be
possible for linearized gravity and for the sum over histories defining
the ground state of isolated systems\refto{19}.  We shall presume that the
analogous contour exists for closed cosmologies although this has yet to
be demonstrated (Problem 3).  The complex nature of the contour in the proposal
(3.23) is not an inessential technicality.  Were the contour purely real,
the wave function would be positive and never oscillate.  With a complex
contour we expect oscillation in some regions of configuration space, and,
as we shall see in Section 4, only in such regions does the wave function
predict the correlations of classical physics.

\section{The Limit of Classical Geometry and Quantum Field Theory in Curved
Spacetime}

In the context of quantum mechanics the predictions of classical physics
are predictions of special kinds of correlations between special classes
of observables.  For example, if we measure the position and momentum of
a particle at one time with accuracies consistent with the uncertainty
principle and then again at a later time the laws of classical physics predict
a definite correlation between these two measurements.

In Section 2 we saw how a wave function predicts correlations among the
observables of an individual system.  It is a very special situation
when the predicted correlations of some observables are classical, but also
a very important situation.  There are three reasons:  First, fundamentally
we interpret the world in classical terms.  Second, certainly in cosmology
our crude observations are of classical observables.  Third, as we shall
describe below, it is possible to give a simple criterion -- the validity
of the semiclassical approximation -- for when a wave function predicts
classical correlations.  This feature greatly simplifies extracting predictions
in quantum cosmology.

The discussion of this section is an attempt at one synthesis of ideas which
have had a long history in general relativity.  Some notable contributions
have been those of Salecker and Wigner\refto{11}, DeWitt\refto{12},
Wheeler\refto{25}, Peres\refto{13}, 
Page and Wooters\refto{14}, Banks\refto{21}, Hawking and Halliwell\refto{22},
 D'Eath and Halliwell\refto{23}, and
Brout, Horwitz and Weil\refto{24}. 

\subsection{The Semiclassical Approximation to Non-Relativistic Particle
Quantum Mechanics}

Let us recall the semiclassical approximation to the wave function of a
state of definite energy in non-relativistic particle quantum mechanics.
 Assuming that the Hamiltonian has been normalized so that the energy is
zero, we want to solve
$$H\psi(X) = 0,\eqno(4.1)$$
where
$$H = - \frac{\hbar^2}{2M} \frac{d^2}{dX^2} + V(X).\eqno(4.2)$$
To obtain the semiclassical approximation we write
$$\psi(X) = \exp \Bigl [iS(X)/\hbar \Bigr ],\eqno(4.3)$$
and expand $S(X)$ in powers of $\hbar$
$$S = S_0 + \hbar S_1 + \cdot\cdot\cdot \eqno(4.4)$$
Writing out the Schr\"odinger equation to the lowest order, $\hbar^0$, one
finds
$$\frac{1}{2M}\Biggl (\frac{dS_0}{dX}\Biggr )^2 + V(X) = 0,\eqno(4.5)$$
so that $S_0$ obeys the classical Hamilton Jacobi equation and, indeed,
is given by \\ $\pm \int^x \sqrt{-2mV(X)}dX$.
In regions where $V(X)<0$ (the ``classically allowed'' regions for $E =
0$) there is a real solution for $S_0$ and the wave function 
oscillates.  In the classically
forbidden regions $S_0$ must be complex, $S_0 = iI_0,$ where $I_0$ solves
the ``Euclidean'' Hamiltonian Jacobi equation.  The wave function in these
regions is a sum of real exponentials.

The next order                                          is also easy to
compute.  The order $\hbar$ part of the Schr\"odinger equation is
$$i \frac{d^2S_0}{dX^2} -2 \frac{dS_0}{dX}\frac{dS_1}{dX} = 0,\eqno(4.6)$$
which is easily solved for $S_1$.  The result, for example in the classically
allowed region is
$$\psi = \Biggl (\frac{dS_0}{dX} \Biggr )^{-{1\over 2}}\exp (iS_0/\hbar).
\eqno(4.7)$$
 In the classically forbidden region the order $\hbar^0$ approximation is
modified by a prefactor in the same way.  Approximate solutions satisfying
given boundary conditions are built by taking linear combinations of (4.7)
and its complex conjugate and of the two possible exponential behaviors
and matching them across the boundaries between the classically allowed
and classically forbidden regions.

In a classically allowed region the interpretation of the semiclassical
approximation is straightforward.  Suppose measurements of the particle's
position and momentum are made with accuracies consistent with the uncertainty
principle.  For position measurements which yield the value $X_0$ the wave
function (4.7) is sharply peaked about the momentum
$$p(X_0) = \Biggl ( \frac{dS_0}{dX} \Biggr )_{X = X_0}.\eqno(4.8)$$ 
This is because near $X_0$
$$S_0(X) \approx S_0 (X_0) + \Biggl( \frac{dS_0}{dX} \Biggr )_{X = X_0}
(X - X_0) + \cdot\cdot\cdot\eqno(4.9)$$
with the higher terms being negligible because $S_0$ is slowly varying.
 In this approximation, (4.7) is a wave function of definite momentum (4.8).
 Thus, a semiclassical approximation of the form (4.7) predicts this classical
correlation between position and momentum.  In particular, if successive
measurements are made which do not disburb either position or momentum
by a large amount, values must be found which are consistent with the classical
equations of motion
$$M \frac {dX}{d\tau} = p(X) = \Bigl [ -2MV(X) \Bigr ]^{1\over 2}.\eqno(4.10)$$
In this way classical physics is recovered.

One should stress that classical physics is recovered only in the sense
of certain correlations and that the nature of these correlations depends
on the form of the semiclassical approximation.  For example, the wave function
(4.7) does not predict much about the position of the particle but only
the correlation (4.8) between position and momentum, and that implied by
(4.10) between present position and future position.  Of course, there are
wave packet states in which
both position and momentum would be predicted, but these do not have definite
energy.  A semiclassical wave function
of the form 
$$\psi (X) = \Biggl ( \frac{dS_0}{dX}\Biggr )^{-{1\over 2}} \cos \Bigl [
S_0(X)/\hbar \Bigr ],\eqno(4.11)$$
would not even predict a correlation of position with momentum but only
with its absolute value.  That is, it would predict that three successive
measurements of position would be correlated according to the equation of
motion (4.10) with the sign of $p(X)$ determined from the first two.

\subsection{The Born-Oppenheimer Approximation for Real Clocks}

In non-relativistic quantum mechanics time is an external parameter labeling
different measurements.  It is not itself an observable.  (See, for example,
the preceding discussion of correlations in the semiclassical approximation.)
 We have access to this time only through the observation of correlations
between the positions of clock indicators and the variables of the system.
 It should, therefore, be sufficient for prediction to formulate quantum
mechanics entirely in terms of these variables.  In the case of the quantum
mechanics of a particle we would write
$$\psi = \psi (T,X),\eqno(4.12)$$ 
where $X$ is the particle's position and $T$ the position of a clock indicator.
We could then study $\psi$ for the correlations between $T$ and $X$ or between
$T$ and other observations.

The wave function in such a formulation of quantum mechanics will satisfy
a constraint reflecting invariance under the choice of parameter used to label
the histories of $T$ and $X$.  Indeed, the parametrized time non-relativistic
quantum mechanics of Section 3.3 is a model for this kind
of theory.  The constraint (3.15) of that theory was
$$H\psi = \Biggl ( -i \frac{\partial}{\partial T} + h \Biggr ) \psi =
 0.\eqno(4.13)$$ 
If we read this as the requirement that the total Hamiltonian vanish,
the variable $T$ may be thought of as the position of a kind of ideal clock
whose Hamiltonian is
$$h_C = -i \frac{\partial}{\partial T} = P_T.\eqno(4.14)$$   
The Hamiltonian (4.14) is a rather poor model of a real clock.  Among other
unrealistic features, its spectrum is unbounded below.  The Hamiltonians of more realistic
clocks we expect to be quadratic in their momenta.  A particle moving freely
in one dimension with narrow dispersions in position and momentum is a simple
 example.
 The position of the particle is a measure of time.  In such a theory, as
we shall show below, we recover the classical notion of time only in the
approximation in which the dynamics of the clock can be treated
semiclassically.

The constraints of general relativity are also quadratic in the momenta.
 In the case of closed cosmologies it does not even seem possible to
approximate the notion of an ideal clock\refto{25}.  Here too we shall recover
a notion of time in the quantum theory only in the approximation
in which spacetime is treated semiclassically.  We shall discuss spacetime
in the next subsection.  Here, we begin with a simple model of a real clock
due to Banks\refto{21,24} which illustrates the central features of these ideas.

Consider a system consisting of a particle described by a position $X$ and
a clock with an indicator variable $T$.  We consider a constraint of the
form
$$H\Psi = (h_C + h)\psi = 0,\eqno(4.15)$$      
where $h$ is the Hamiltonian of the particle and $h_C$ the Hamiltonian of
the clock.  The dynamics of the clock we take to be specified by the action
$$S[T(\tau)] = M \int d\tau \Bigl (\frac12 \dot T^2 + V_C (T) \Bigr ),
\eqno(4.16)$$ 
so that $h_C$ will be quadratic in the momenta.  The quantity $M$ is the
mass of the clock although it also controls the coupling to the potential
$V_C$.

In the limit $M \rightarrow \infty$ the clock motion can be treated
classically.  Physically, this is because, for given energy, as the mass
becomes large the quantum fluctuations become small.  Mathematically, it
is because the classical limit in a sum over histories occurs as $\hbar
\rightarrow 0$ in $\exp (iS/\hbar)$ leading to destructive interference in
the sum for all but the classical trajectory.  The limit $M \rightarrow
\infty$ is the same limit but for the clock part of the action alone.  The
approximation of large $M$ is therefore not strictly the semiclassical
approximation.  Rather it is the analog of the Born-Oppenheimer approximation
in molecular physics in which the motion of the massive nuclei are considered
classically while the electronic cloud is treated quantum mechanically.

In the large $M$ limit we look for a solution of the constraint equation
(4.15) of the form
$$\psi = e^{iS(T)}\chi(T,X)\eqno(4.17)$$
where
$$S = MS_0 + S_1 + M^{-1}S_2 + \cdot\cdot\cdot\eqno(4.18a)$$  
$$\chi = \chi_0 + M^{-1}\chi_1 + \cdot\cdot\cdot\eqno(4.18b)$$  
Since
$$h_C = -\frac{1}{2M}~\frac{d^2}{dT^2} + MV_C (T),\eqno(4.19)$$   
we have, on writing out the constraint (4.15),
$$           \Biggl \{  - \frac{1}{2M} \Biggl [ i \frac{d^2S}{dT^2} - \Biggl
( \frac{dS}{dT} \Biggr )^2 \Biggr ] + MV_C \Biggr \} \chi   
 - \frac{1}{2M}~\frac{\partial^2\chi}{\partial T^2} - \frac{i}{M}~
\frac{dS}{dT}~\frac{\partial\chi}
{\partial T} + h\chi = 0.\eqno(4.20)    $$ 

We now insert the expansions (4.18) in (4.20) and systematically expand in powers
of $M$.  For the leading order we find
$$\frac12 \Bigl ( \frac{dS_0}{dT} \Bigr )^2 + V_C (T) = 0.\eqno(4.21a)$$
This is the clock's classical Hamilton-Jacobi equation.  Defining 
$p_T = M(dS_0/dT)$ it can be written in the familiar form
$$\frac{1}{2M} p^2_T + MV_C(T) = 0.\eqno(4.21b)$$
In the next order, $M^0$, one finds
$$-\frac12 \Biggl ( i \frac{dS_0}{dT^2} - 2 \frac{dS_0}{dT}~\frac{dS_1}{dT}
\Biggr ) \chi_0 - i \frac{dS_0}{dT}~\frac{\partial\chi_0}{\partial T} +
 h\chi_0 = 0.
\eqno(4.22)$$                                                              
This is one equation for two unknowns.  To fix the remaining freedom we
rewrite (4.22) by {\it defining} a classical time from the
solution to the classical Hamilton-Jacobi equation (4.21).  We write
$$M \frac{dT_0}{d\tau}= p_T = M \frac{dS_0}{dT}.\eqno(4.23)$$   
Integration of this relation defines $T_0(\tau)$ up to an initial condition
and hence $\tau$ as a function of $T$.
  Equation (4.20) can now be rewritten to read
$$-\frac12 \Biggl ( i \frac{d^2S_0}{dT^2} - 2 \frac{dS_1}{d\tau}       
\Biggr ) \chi_0 - i \frac{\partial\chi_0}{\partial\tau} + h\chi_0 = 0,
\eqno(4.24)$$                                                              
where $\chi_0 = \chi_0 (T_0(\tau),X)\equiv \chi_0(\tau, X).$  An inner product
can be defined 
by integrating over $X$ at constant $\tau$
$$(\chi, \phi) = \int dX~\chi^*(\tau, X) \phi (\tau, X)\eqno(4.25)$$  
and in it we can take the expectation value of (4.24).  The result is
$$-\frac12 \Biggl ( i \frac{d^2S_0}{dT^2} - 2 \frac{dS_1}{d\tau}       
\Biggr ) (\chi_0, \chi_0) - i \Biggl ( \chi_0, \frac{\partial \chi_0}
{\partial \tau} \Biggr ) + (\chi_0, h \chi_0) = 0.\eqno(4.26)$$         
The imaginary part of this equation is, assuming that $h$ is Hermitian,
$$\frac12 \Biggl ( - \frac{dS_0}{dT^2} + 2  \frac{d Im S_1}{d \tau}       
\Biggr ) (\chi_0, \chi_0) - \frac12~\frac{d}{d\tau} (\chi_0, \chi_0) = 0.
\eqno(4.27)$$                                                            
The real part is
$$\frac{d(Re S_1)}{d\tau} (\chi_0, \chi_0) - \frac{i}{2} \Biggl [ \Biggl(
\chi_0, \frac{\partial \chi_0}{\partial \tau} \Biggr ) - \Biggl( \frac{
\partial \chi_0}{\partial \tau}, \chi_0 \Biggr ) \Biggr ] + (\chi_0, h
\chi_0) = 0.\eqno(4.28)$$
We recover a sensible quantum mechanics if we impose the condition that
the inner product is conserved
$$\frac{d}{d\tau} (\chi_0, \chi_0) = 0\eqno(4.29)$$
so that from (4.27)
$$- \frac{dS_0}{dT^2} + 2 \frac{dS_0}{dT}~\frac{d Im S_1}{dT} = 0.\eqno(4.30)$$
This is the usual next -after-leading-order equation for the semiclassical
approximation to the $T$ motion.  Equation (4.26) becomes, assuming $\chi_0$
normalized,
$$\frac{d(Re S_1)}{d\tau} + (\chi_0, h \chi_0) = 0 .\eqno(4.31)$$ 
When (4.30) and (4.31) are substituted back into the original equation (4.22)
we find
$$i \frac{\partial \chi_0}{\partial\tau} = \Bigl [h - (\chi_0, h \chi_0)
\Bigr ] \chi_0.\eqno(4.32)$$ 
This is the Schr\"odinger equation for the particle moving in the ``background''
time $\tau$ to the extent $(\chi_0, h \chi_0)$ is constant or negligible.

The combined equation for $Re S$ accurate to first order in $M$ is from
(4.31) and (4.21)
$$\frac12 \Biggl ( \frac{d Re S}{dT} \Biggr )^2 + MV_C(T) + (\chi_0, h\chi_0)
= 0.\eqno(4.33)$$
This is the classical Hamilton-Jacobi equation but with a small quantum
correction to the energy.
 It is the semiclassical {\it back reaction} equation.

\subsection{The Approximation of Quantum Field Theory in Curved Spacetime.}

The structure of the Hamiltonian constraint for general relativity -- the
Wheeler-DeWitt equation -- is similar both in structure and origin to the
constraint of the simple model just discussed.  Including the energy of
a scalar matter field, the Wheeler-DeWitt equation reads
$$\Biggl [ - \frac12 \ell^2 \nabla^2_x + \frac12 \ell^{-2}h^{1\over 2} (
2 \Lambda - ^3 R) + h^{1\over 2} T_{nn} \Biggl ( \Phi, - i \frac{\delta}
{\delta   \Phi} \Biggr ) \Biggr ] \Psi [h_{ij}, \Phi ] = 0.\eqno
(4.34)$$
Here $T_{nn}(\Phi, \Pi)$ is the stress-energy of the matter field expressed
in terms of the field's value and momentum and projected onto the normals
of the spacelike hypersurface.  It is the Hamiltonian density
for the scalar field.  The inverse squared Planck length enters
the constraint in exactly the same way as the mass of the clock did in the
model problem [cf. (4.15),(4.19)].  We may therefore
consider the limit when $\ell \rightarrow 0$ and expect to treat geometry
semiclassically\refto{21-24}.  This is the limit when relevant length scales 
are large
compared to the Planck length and when relevant energies are small compared
to the Planck mass.

The model of Section 4.2 reveals the central features of the $\ell \rightarrow
0$ limit so clearly that we shall just sketch the parallel steps here. 
We write
$$\Psi [h_{ij}, \Phi ] = e^{iS      [h_{ij}       ]} \chi
      [h_{ij}, \Phi       ] \eqno(4.35)$$
and systematically expand $S, \chi$ and the Wheeler-DeWitt equation in powers
of $\ell$.
In the leading order we recover the Hamilton-Jacobi equation of general
relativity for $S_0$.  From its solution we can introduce the momenta
$$\ell^2 \pi^{ij}({\bf x}) = \frac{\delta S_0      [h_{ij}({\bf x})      ]}
{\delta h_{ij}({\bf x})}.\eqno(4.36)$$ 
The $\pi_{ij}$ are the tangent vectors to a set of integral curves in
superspace which are solutions to the classical Einstein equation.  For
example, if we work in a gauge in which 4-metrics have the form
$$ds^2 = -d\tau^2 + h_{ij} (\tau,{\bf x}) d x^i dx^j \eqno(4.37)$$
then classically [cf. (3.17)]
$$\pi_{ij} = \frac{1}{2\ell^2 h^{1\over 2}}~\frac{d}{d\tau}(hh_{ij})
.\eqno(4.38)$$ 
Integrating
$$\frac{dh_{ij}}{d\tau} = G_{ijk\ell}\frac{\delta S_0}
{\delta h_{k\ell}},\eqno(4.39)$$
we recover a time dependent 4-geometry in the gauge (4.37) which satisfies
the Einstein equation.

The values of $\chi_0$ along an integral curve in superspace define $\chi_0$ 
as a function
of $\tau$.
$$\chi_0 = \chi_0       [ h_{ij} (\tau,{\bf x}), \Phi({\bf x})      ]
 = \chi_0       [\tau, \Phi({\bf x})      ].\eqno(4.40)$$ 
Then by exactly the same steps as led from (4.17) to (4.32) one finds starting
from (4.35) that one ends at
$$i \frac{\partial \chi_0}{\partial \tau} = \Biggl [ h^{1\over 2}T_{nn}
 \Biggl ( \Phi, -
i \frac{\delta}{\delta \Phi} \Biggr ) - \Biggl (\chi_0, h^{1\over 2}T_{nn}
 \chi_0
\Biggr ) \Biggr ] \chi_0.\eqno(4.41)$$
This is a Schr\"odinger equation defining a quantum field theory in the
curved background spacetime specified by a solution to (4.39).  $T_{nn}$
is the Hamiltonian density for the matter field which depends on the background
metric in the usual way.  The inner product with which this equation is
derived is the standard inner product in the field representation on the
spacelike surface with normal $\partial/\partial \tau$ in the background
(4.37), i.e.
$$(\chi, \chi^{\prim}) = \int \delta \Phi \chi^* [\tau, \Phi ({\bf x})]
\chi^{\prim} [\tau, \Phi ({\bf x})].\eqno(4.42)$$

Accurate to order $\ell^2,$ the operator constraints of general relativity
imply the classical constraint equations corrected by the quantum expectation
value of the stress-energy of the matter field.  The derivation is parallel
to that of (4.33).  In the Hamilton-Jacobi form in which they naturally
emerge from (4.32) and (4.33) one has
$$\frac12 \Biggl [ - \ell^4 G_{ijk\ell} \frac{\delta S}{\delta h_{ij}}~\frac
{\delta S}{\delta h_{k\ell}} + h^{1\over 2}(^3 R - 2\Lambda)\Biggr ] = 
\frac{\ell^2}{2}~h^{1\over 2} \Bigl (\chi_0, T_{nn} \chi_0 \Bigr ),
\eqno(4.43a)$$
$$D_j \Biggl ( \frac{\delta S}{\delta h_{ij}} \Biggr ) =
 \frac{\ell^2}{2}~h^{1\over
2} \Bigl (\chi_0, T^i_n \chi_0 \Bigr ),\eqno(4.43b)$$ 
or equivalently using eqs. (4.36) - (4.39)
$$\Big ( R^{\alpha}_n - \frac12 \delta^{\alpha}_n R \Bigr ) = \frac{\ell^2}{2}
\Big (\chi_0, T^{\alpha}_n \chi_0 \Big ).\eqno(4.44)$$
where $R_{\alpha\beta}$ and $T_{\alpha\beta}$ are the Ricci curvature 
and stress energy expressed
in an orthonormal basis one
member of which is $n^{\alpha}$.  These are the four constraint equations
of the classical theory including the quantum corrections to the stress
energy of the matter field.

The correspondence in form between (4.43) and (4.44) shows that any solution
of the former will satisfy four of the quantum corrected Einstein equations.
 However, these Hamilton-Jacobi equations
determine  a solution of the {\it full} set of Einstein equations through
(4.36) - (4.39).   The reason is covariance.  There was no intrinsic definition
of the spacelike surface with normal $n^{\alpha}$.  The four constraint
equations (4.44) must be satisfied on {\it any} spacelike surface in the
 geometry.
 This requirement is equivalent to the full set of Einstein equations\refto
{26}.
Thus,
$$R_{\alpha\beta} - \frac12 g_{\alpha\beta} R = \frac{\ell^2}{2}(\chi_0, T_{\alpha
\beta} \chi_0).\eqno(4.45)$$
These are the quantum corrected (``backreaction'') Einstein equations for
the background geometry.  (Problem 4).

The problem of extracting the predicted correlations from the wave function
of the universe is in general a difficult one.  As the results of this section
argue, however, if the wave is well approximated by the form (4.35), with
$\chi$ and $S$ given by the first few terms in an expansion in powers of
the Planck length, then the correlations of a classical 4-geometry containing
quantum matter fields are predicted.  The 4-geometry and quantum fields
are defined precisely through equations (4.36 - 4.39) and (4.40 - 4.41).
In the context of the discussion of
Section 2 we envision this means the following:  Imagine filling space with
a system of rods, clocks and field meters which could define what is meant
by measurements of distance and time to classical accuracies and measurements
of field.  The clocks, rods and meters are to be described by matter and
gravitational fields so that it is possible to identify the regions in
superspace consistent with various possible sets of values they might read.
 From the analog with non-relativistic quantum mechanics, we expect that,
where the semiclassical approximation is valid, the wave function will
be sharply peaked about a region consistent with values which define
 spacetimes satisfying the Einstein equations (4.45) and quantum fields
satisfying (4.41).  

In quantum cosmology, an important test of any theory of initial conditions
is that there be a region of configuration space in which the wave function
is well approximated semiclassically for we observe the present universe
to behave classically.  However,the wave function in the semiclassical 
approximation does not predict a unique classical history.  Rather, it predicts
a family of them.  For example, in particle quantum mechanics a semiclassical
wave function of the form $\exp [iS(X)/\hbar]$ corresponds to classical
trajectories with momentum $p = dS/dX$.  Information about present position
must be used to single out a unique classical history.  A semiclassical
wave function of the form $\cos [S(X)/\hbar]$ requires even more information
for it corresponds to classical trajectories with $p = \pm dS/dX$.  How
much present information must be used to gain definite, classical predictions
from the wave function of the universe is one of the subject's most important
questions.

\subsection{The Semiclassical Vacuum}

The derivation of quantum field theory in curved spacetime presented in
the preceding subsection clarifies a number of issues usually regarded as
internal to that subject.  For example, the derivation sheds light on the
meaning of the metric in the semiclassical field equation (4.45).  It emerges
there not as the expectation value of a field operator in some {\it general}
 quantum
state.  Rather, $g_{\alpha\beta}$ is the metric that would be determined
from {\it classical measurements of limited occuracy} in a regime of 
configuration space in which the semiclassical approximation to a {\it 
particular} wave function is valid.

As a second example, the derivation of quantum field theory in curved
spacetime connects the choice of ``vacuum'' for
the matter
with a theory of initial conditions i.e. with a
prescription for the wave function of the universe.  If the semiclassical
approximation is valid, $\chi_0 [\tau, \Phi ({\bf x})]$ is determined by
$\Psi$ through (4.35), (4.40) and this functional defines a quantum state
of the matter field in the classical background spacetime in the Hilbert
space defined by (4.42).  We shall call it the quantum state of the matter
fields.

The determination of the quantum state of the matter fields by a prescription
for the wave function of the universe may be instructively illustrated in
a simple minisuperspace model.  For the prescription we consider that of
 Section 3.3.  For the minisuperspace model we restrict attention to homogeneous
and isotropic geometries containing a single {\it conformally invariant}
scalar field.  This model is easy to analyse even it it is not very realistic.

The metric of a homogeneous isotropic geometry can be put in the form
$$ds^2 = \sigma^2 \Bigl [ - d\tau^2 + a^2(\tau)d\Omega^2_3 \Bigr ]\eqno(4.46a)$$
$$= \sigma^2 a^2(\eta) \Bigl [-d\eta^2 + d\Omega^2_3 \Bigr ],\eqno(4.46b)$$
where $\sigma^2 = \ell^2/24 \pi^2$ is a convenient normalizing constant.
The geometry of a three surface of constant $\tau$ is then characterized solely
by its radius $a_0$.  The wave function is a {\it function} of $a_0$ and
a {\it functional} of the matter field on this surface.
$$\Psi = \Psi \Bigl [a_0, \Phi_0 ({\bf x}) \Bigr ].\eqno(4.47)$$
Equivalently, we could expand any field configuration in harmonics
$$\Phi (\eta, {\bf x}) = \sum_n \Phi^{(n)} (\eta) Y_{(n)} ({\bf x}),\eqno(4.48)
$$
where the $Y_{(n)}({\bf x})$ are standard harmonics on the 3-sphere.  Then
$$\Psi = \Psi \left[a_0,
\Phi^{(1)}_0,\Phi^{(2)}_0,\Phi^{(3)}_0,\cdot\cdot\cdot\right ].\eqno(4.49)$$

The action is the sum of the Euclidean gravitational action (3.24)
and the Euclidean action
for the scalar field corresponding to (3.2) with $\xi = 1/6$.  The Euclidean
 gravitational
action restricted to the minisuperspace geometries is
$$I_E = \frac12 \int d\eta \Bigl [ -(a^{\prim})^2 - a^2 + H^2 a^4 \Bigr
],\eqno(4.50)$$
where a prime denotes an $\eta$-derivative and $H^2 = \Lambda/3$.  The matter
action is considerably simplified by a dimensional rescaling
$$\Phi (x) = \varphi(x) / (2\pi^2\sigma^2)^{1\over 2}\eqno(4.51)$$
and by a conformal rescaling
$$\varphi^{(n)}(\eta) = \chi^{(n)}(\eta) /a(\eta).\eqno(4.52)$$ 
The physics of the field, being conformally invariant,
is essentially, the same in all conformally related spacetimes.
The geometry is conformal to an Einstein static universe by a conformal
factor $a(\eta)$.  In that geometry, because of its timelike killing field,
the analysis of the scalar field is considerably simplified.  This is 
immediately apparent in the form of the matter action takes in terms of
the variables $\chi^{(n)}$
$$I_M = \frac12 \int d\eta \Biggl [ \Bigl (\chi^{(n){^\prim}}\Bigr )^2 +
\omega^2_n \Bigl (\chi^{(n)} \Bigr )^2 \Biggr ],\eqno(4.53)$$
where $\omega^2_n = \gamma_n+1$ and $\gamma_n$ are the eigenvalues of
the Laplacian on the 3-sphere.

In the minisuperspace model the wave function is given as
$$\Psi \Bigl [a_0,\chi^{(n)}_0 \Bigr ] = \int \delta a \prod_n \delta \chi^{(n)}
e^{- \left (I_E + I_M \right )}.\eqno(4.54)$$
The integral is over all $a(\eta)$ which correspond to compact geometries
with boundary three sphere radius $a_0$, and over all matter mode
configurations $\chi^{(n)}(\eta)$ which match $\chi^{(n)}_0$ on the boundary
and are elsewhere regular. (Figure 3).  A compact geometry
 will have
one radius (the ``south pole'') at which $a$ vanishes linearly in the polar
angle $\tau$.  Since $d\eta = d\tau /a(\eta)$, the coordinate $\eta$ becomes
logarithmically infinite at this point.  We may use the last remaining gauge
freedom to choose the boundary to be at $\eta = 0$.  The relevant co\"ordinate
range for $\eta$ is then $(-\infty, 0)$.  We thus integrate over $a(\eta),
 \chi(\eta)$ which vanish at $\eta = -\infty$ and assume the prescribed
values on the boundary.

\begin{figure}[t!] 
\centerline{\includegraphics[height=3in]{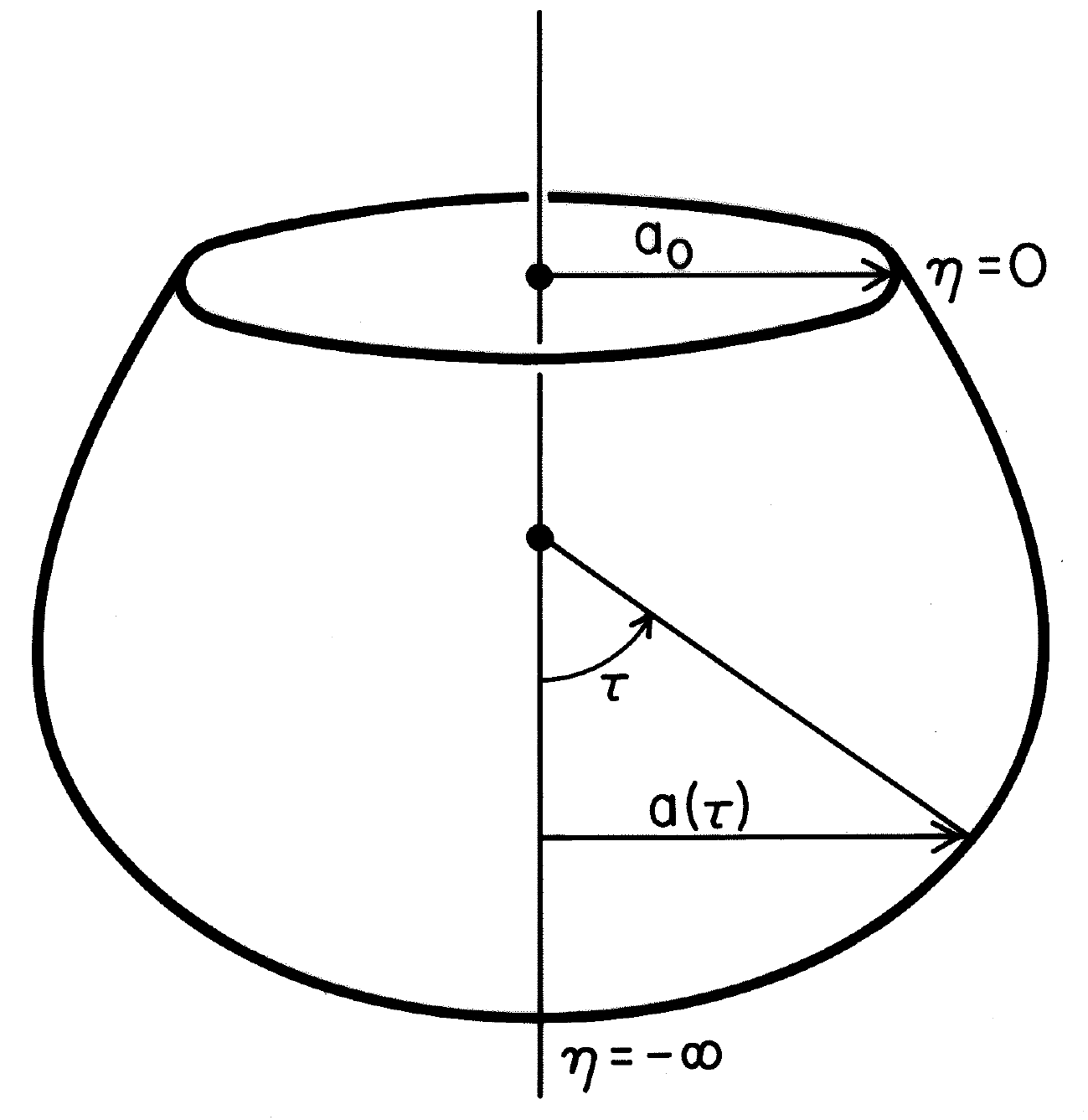}}
\caption{  A two dimensional representation of a homogeneous and
isotropic 4-geometry contributing to the sum for the state of minimum
excitation $\Psi (a_0, \Phi_0)$.  Shown embedded in a flat 3-dimensional
space is a 2-dimensional slice of such a geometry whose intrinsic geometry
is
%$$d\sigma^2 = d\tau^2 + a^2 (\tau) d \varphi^2$$
$\tau$ is thus a ``polar angle'' and $a$ the ``radius from the axis.'' 
The geometry is compact and has only one boundary at which the radius is
$a_0$, the argument of $\Psi$.  The field configurations $\Phi (\tau, {\bf
x})$ which contribute to the sum are those which are regular on this surface
and which match the argument of the wave function $\Psi$ on the boundary.} 
\end{figure}

\begin{figure}[t!]
\centerline{\includegraphics[height=3in]{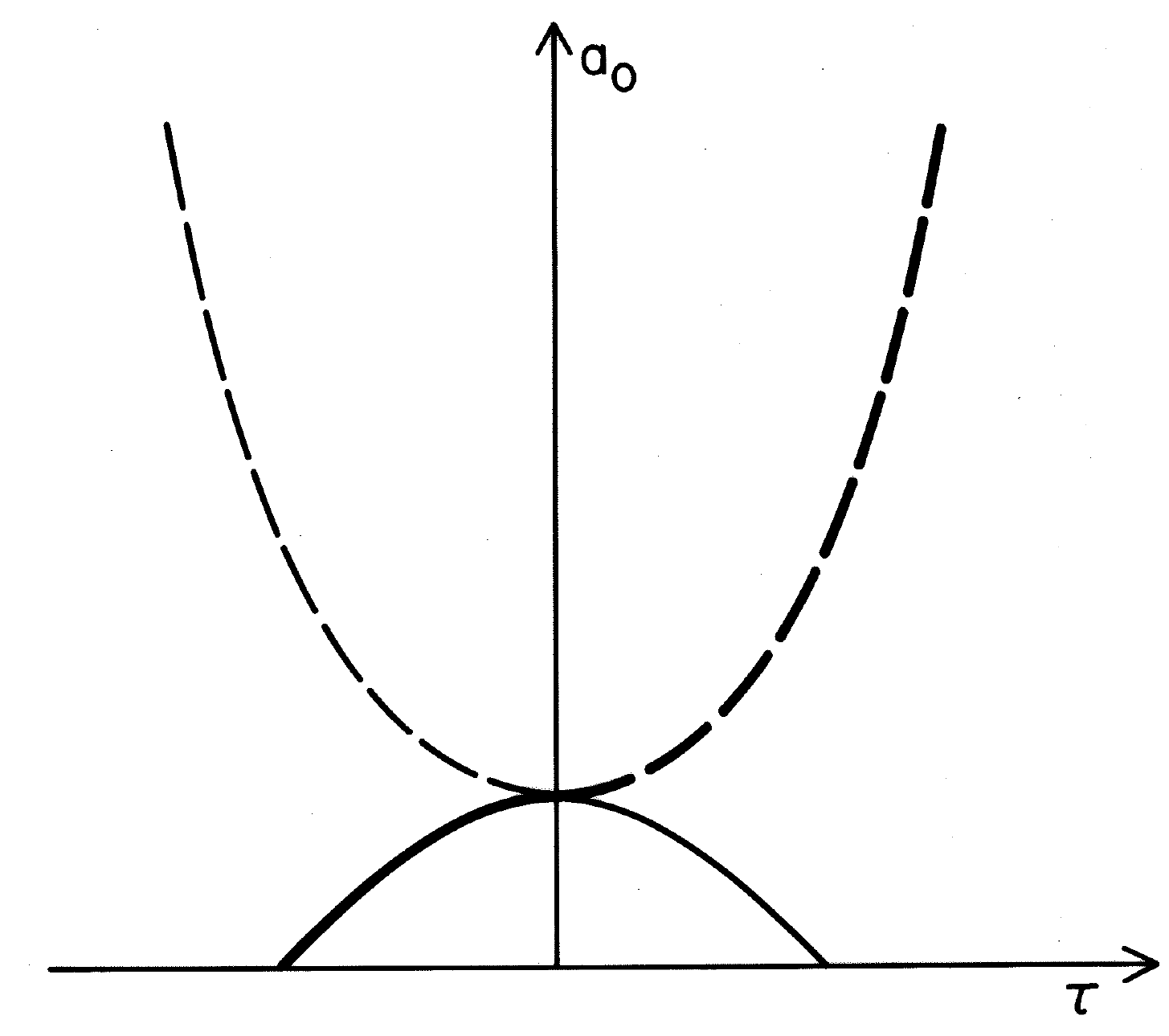}}
\caption{The extremizing scale factor for the homogeneous, isotropic
minisuperspace model with conformally invariant scalar field.  The solid
line is the solution of (4.58) for real Euclidean extrema of the action.
 The complete range of $a$ from zero to maximum and back again describes the
geometry of the 4-sphere (Figure 5).  The dashed curve is the solution of
(4.60) for complex Euclidean (Lorentzian) extrema.  It describes the geometry
of de Sitter space.  For each value of $a_0$ there are thus two possible
extremizing solutions.  Choosing the trajectory to start on the left at
$a_0 = 0$, the Euclidean prescription for the ground state singles out the
heavy curve shown.  This gives the semiclassical approximation to the wave
function $\Psi$.}
\end{figure}                                                               

\begin{figure}[t!]
\centerline{\includegraphics[height=3in]{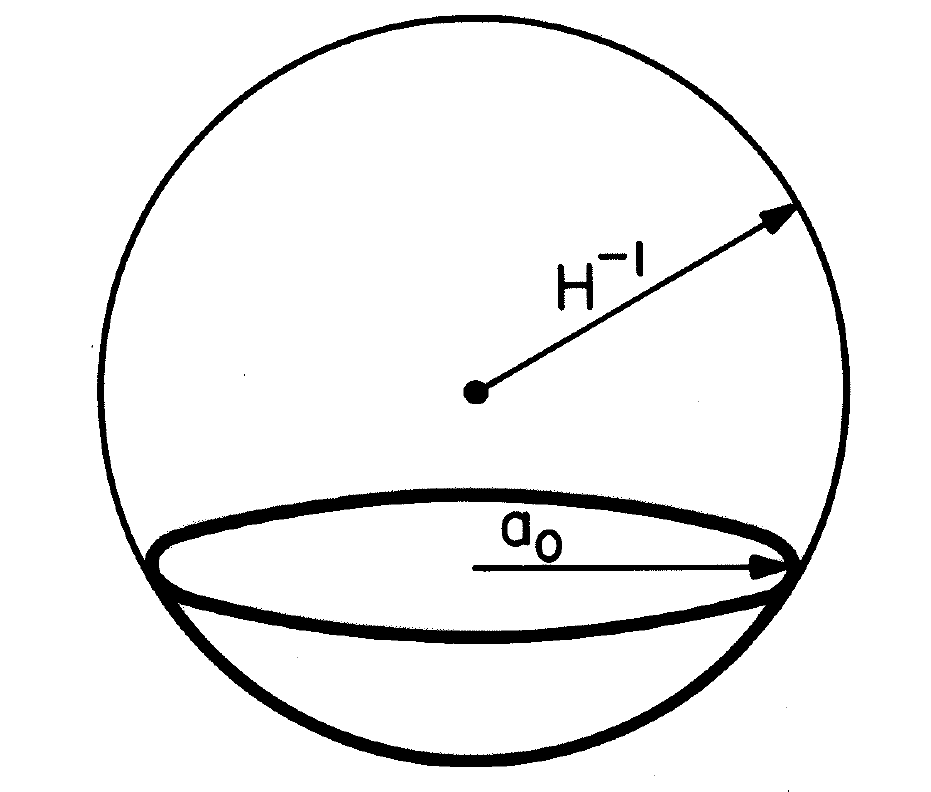}}
\caption{  The real Euclidean extrema of the homogeneous, isotropic
minisuperspace model with conformally invariant scalar field have the geometry
of a 4-sphere of radius $H^{-1}$.  The extremizing configuration which gives
the semiclassical approximation to $\psi$ at $a_0 < 1/H$ is a part of the
4-sphere with a single 3-sphere boundary of radius $a_0$.  There are two
possibilities corresponding to more than a hemisphere or less.  The Euclidean
functional integral prescription for $\psi$ identifies the smaller part
of the 4-sphere as the contributing extremum.  For $a_0 > H^{-1}$ there are
no real extrema.  The orientation of the 3-sphere in the 4-sphere is
arbitrary.  The semiclassical vacuum of the matter field is thus de Sitter
invariant.  }
\end{figure}                                                              

In the simplicity of conformal invariance, the action separates into a sum
of metric part and field part so that the two integrals can be done
 separately.  The integral over
the field part is trivial.  One finds
$$\Psi \Bigl [ a_0, \chi_0^{(n)} \Bigr ] = \psi (a_0) \prod_n \exp \Biggl [
- \frac12 \omega_n \Bigl (\chi^{(n)}_0 \Bigr )^2 \Biggr ],\eqno(4.55)$$
where
$$\psi (a_0) = \int \delta a e^{-I_E[a]}.\eqno(4.56)$$
Let us now approximate the integral for $\psi(a_0)$          
by the method of steepest descents.  For this we must find the extrema of
$I_E$ through which the contour of integration 
can be distorted.  We begin
with values of $a_0$ less than $H^{-1}$.  The possible extrema of $I_E$ are
just the solutions of
$$a^{\prim\prim} - a + 2H^2a^3 = 0.\eqno(4.57)$$
The equation has an ``energy integral'' whose value may be found from the
regular vanishing of the $a$ at $\eta = -\infty$.  Expressing this integral
in terms of $\tau$ gives
$$\Bigl ( \frac{\dot a}{a}\Bigr ) ^2 = \frac{1}{a^2} - H^2.\eqno(4.58)$$
This is the Euclidean Einstein equation for a metric with the symmetries
of the model as it must be.  The solution is illustrated in Figures 4 and
5 and is just the 4-sphere of radius $1/H$.  For $a_0 < 1/H$ there are
thus two possible extrema which are compact 4-geometries with a 3-sphere
boundary of radius $a_0$.  One for which the boundary bounds less than a
hemisphere of the 4-sphere and another for which it bounds more.  The action
for the 4-sphere is negative and therefore one might think that the extremum
encompassing more 4-sphere should dominate.  One must remember, however,
that because of the conformal rotation the contour of $a$ integration is
in the imaginary direction in the immediate vicinity of the extremum.  
Extrema of analytic functions are saddle points so that a maximum in a real
direction is a minimum in an imaginary direction.  The stationary configuration
which contributes to the steepest descent evaluation of (4.56) is the one
which is a maximum of the action in real directions and a least action
configuration in imaginary directions.  The extremum corresponding to the
smaller part of the 4-sphere, therefore, provides the steepest descent
approximation to the wave function.  In fact, the contour cannot be distorted
to pass through the other extremum.  We thus have for $a_0 < 1/H$
$$\psi (a_0)\approx N \Biggl [ -1+a^2_0-H^2a^4_0 \Biggr
 ]^{-1/4}  
\exp \Biggl [ - \frac{1}{3H^2} \Bigl (1-H^2 a^2_0 \Bigr )^{3/2}
\Biggr ],\eqno(4.59)$$ 
where $N$ is an arbitrary normalizing factor.

If $a_0$ is increased to a value larger than $1/H$ there are no longer any
real extrema because a 3-sphere of radius $a_0 > 1/H$ cannot fit into a
4-sphere of radius $1/H$.  There are, however, complex extrema.  These can
be obtained by changing $\tau \rightarrow \pm$ {\it it} in Equation (4.58) so
they solve
$$\Bigl ( \frac{\dot a}{a}\Bigr ) ^2 = H^2 - \frac{1}{a^2}.\eqno(4.60)$$
An extremum is then a solution of the Lorentzian Einstein equations with positive
cosmological constant.  This solution is called de Sitter space 
.  These complex extrema must contribute in complex conjugate pairs so
that the wave function is real.  By a standard WKB matching analysis we
can establish the form of the wave function for $a_0 > H^{-1}$
$$\psi (a_0) \approx 2N \left [H^2a^4_0 - a^2_0 + 1 \right
]^{-1/4}
\cos \Biggl [ \frac{\Bigl ( H^2a^2_0 - 1 \Bigr )^{3/2}}{3H^2} -
\frac{\pi}{4} \Biggr ].\eqno(4.61)$$
This form could be derived by carefully following the extremum configuration
as $a_0$ is increased along the heavy curve shown in Fig. 4.

In the region $a_0 > H^{-1}$ the steepest descents approximation to the
wave function is a real linear combination of solutions of the semiclassical
form (4.33).  In this 
          region we find the geometric correlations of
classical de Sitter space.  The state of the matter field can be read of
(4.35), (4.40) and (4.56).  It is, up to normalization, 
$$\chi_0 \bigl [ \tau, \varphi^{(n)}_0 \bigr ] = \prod_n \exp \Biggr [
- \frac12\omega_{(n)} \Bigl (a(\tau)\varphi^{(n)}_0 \Bigr )^2 \Biggr
].\eqno(4.62)$$ 
In the minisuperspace model, the prescription for the wave function of the 
universe of Section 3.3 predicts this particular quantum
state of matter in the limit of classical geometry and quantum field theory
in curved spacetime.

The state specified in the field representation by (4.62) is familiarly
known as the Euclidean de Sitter invariant vacuum state.  To see this we
note that in terms of the variables $\chi^{(n)}_0$ the wave function (4.62)
is that of a field in its ground state in the Einstein static universe.
 In Fock space the state is therefore annihilated by the modes which are
positive frequency in the conformal time $\eta$.  In terms of $\varphi$
these are modes proportional to
$$[a(\eta)]^{-1} \exp (- i \omega_{(n)} \eta) Y_{(n)} ({\bf x}).\eqno(4.63)$$
This is the conventional definition of the Euclidean, de Sitter invariant
vacuum\refto{26}.

As argued by D'Eath and Halliwell\refto {23} the de Sitter invariance of the quantum
state of the matter is an inevitable consequence of the symmetry of the
Euclidean sum over histories prescription.  In the defining region $a_0
< H^{-1}$ the extremizing configuration which supplies the steepest descents
approximation is the smallest part of a 4-sphere bounded by a 3-sphere of
radius $a_0$.  However, the wave function does not depend on the orientation
of the 3-sphere in the 4-sphere.  There is thus an $O(5)$ invariance which in
the Lorentzian region corresponds to the de Sitter group.

\vskip .3in

Preparation of these lectures was supported in part by the National Science
Foundation under Grant PHY 85-06686.
%\end{acknowledgements}

\appendix
\section{Problems}

\begin{description}
\item {1.}  Practice expressing cosmological and local observations in terms
of correlations in the wave function of the universe.  Are there observations
which cannot be so expressed?

\item {2.}  Provide a careful derivation of the Wheeler-DeWitt equation from the
sum over histories which shows the connection between the factor ordering
and the measure.  Hint:  See Ref. 28.

\item {3.}  Is there a complex contour in the space of 4-geometries and
matter field configurations along which the sum over histories for quantum
cosmology is convergent?  

\item {4.}  Discuss the regularization of quantum field theory in curved
spacetime in the context of the semiclassical limit of quantum cosmology
described in Section 4.  Hint:  See Ref. 23.
\end{description}

\section{Notation}

For the most part we follow the conventions of Ref. 29 with respect to
signature, curvature and indices.  In particular:

\noindent     {\it Signature}:  (-,+,+,+) for Lorentzian spacetimes.
                       (+,+,+,+) for Euclidean spacetimes.

\noindent     {\it Indices}:     Greek indices range over spacetime from 0 to 3.
                       Latin indices range over space from 1 to 3.

\noindent     {\it Units}:         We use units in which $\hbar = c = 1$.  The Planck
                       length is $\ell = (16 \pi G)^{1\over 2} = 1.15 \times
10^{-32}$ cm.

\noindent     {\it Covariant Derivatives}:  $\nabla_\alpha$ denotes a spacetime
 covariant derivative and $D_i$ a spatial one.

\noindent     {\it Traces and Determinants}:  Traces of second rank tensors
$K_{\alpha\beta}$ are written as $K = K^\alpha_\alpha$ except when the tensor is
the metric in which case $        g$ is the determinant of $g_{\alpha\beta}$
and $h$ the determinant of $h_{ij}.$  

\noindent     {\it Extrinsic Curvatures}:  If $n_\alpha$ is the unit normal to
a spacelike hypersurface in either a Euclidean or Lorentzian spacetime,
we define its extrinsic curvature to be
$$K_{ij} = \nabla_i n_j.$$

\noindent     {\it Intrinsic Curvatures}:  Intrinsic curvatures are defined so
that the scalar curvature of a sphere is positive.

\noindent     {\it Metric on the unit n-sphere}:  This is denoted by $d\Omega^2_n$
and in standard polar angles is
$$d\Omega^2_2 = d\theta^2 + \sin^2 \theta d \varphi^2~~~~~n = 2$$
$$d\Omega^2_3 = d\chi^2 + \sin^2 \chi d \Omega^2_2~~~~~n = 3$$

\section{References}

\begin{description}
\item {1.}  Discussions of the general problems encountered in the search
for initial conditions may be found in R. Penrose, in {\it General
Relativity:  an Einstein Centenary Survey}, ed. by S. W. Hawking and W. Israel
(Cambridge University Press, Cambridge, 1979), J. B. Hartle in {\it    
Inner Space/Outerspace:  the Interface between Cosmology and Particle Physics},
ed. by E. W. Kolb, et.al. (University of Chicago Press, Chicago, 1986) and
J. Barrow and F. Tipler, {\it The Anthropic Principle}, (Clarendon Press,
Oxford, 1986).  A sample of specific proposals for laws for initial conditions
may be found in the article by Penrose cited above and in Refs 2-5.

\item {2.}  S. W. Hawking, in {\it Astrophysical Cosmology:  Proceedings
of the Study Week on Cosmology and Fundamental Physics} ed. by H. A. Br\"uch,
G. V. Coyne and M. S. Longair (Pontificiae Academiae Scientiarum Scripta
Varia, Vatican City, 1982),   Nucl. Phys., B239, 257, 1984
and J. B. Hartle and S. W. Hawking,  Phys. Rev., D28, 2960, 1983.

\item {3.}  A. Vilenkin,  Phys. Lett., B117, 25, 1983, 
Phys. Rev., D27,
2848, 1983, Phys. Rev., D30, 509, 1984, Phys. Rev., D32,
 2511, 1985, TUTP
 preprint 85-7.

\item {4.}  J. V. Narlikar and T. Padmanabhan, Physics Reports,
100, 151, 1983, T. Padmanabhan ``Quantum Cosmology - The Story So Far''
(unpublished lecture notes).

\item {5.}  W. Fischler, B. Ratra, L. Susskind,  Nucl. Phys. B,
259, 730, 1985.

\item {6.}  The author's lectures at Cargese were somewhat more extensive
than presented here.  Some of this material can be found in the author's
lectures in
{\it High Energy Physics 1985:  Proceedings of the Yale Theoretical 
Advanced Study Institute} ed. by M. J. Bowick and F. G\"ursey (World Scientific,
Singapore 1985).

\item {7.}  For a clear review of the basic ideas in canonical quantum gravity
in greater detail than can be presented here see K. Kuchar, in {\it Relativity
Astrophysics and Cosmology} ed. by W. Israel (D. Reidel, Dordrecht, 1973)
and in {\it Quantum Gravity 2} ed. by C. Isham, R. Penrose and D. Sciama
(Clarendon Press, Oxford, 1981).

\item {8.}  See, e.g., H. Everett,   Rev. Mod. Phys., 29, 454, 1957,
the many articles reprinted and cited in {\it The Many Worlds Interpretation
of Quantum Mechanics}, ed. by B. DeWitt, and N. Graham, (Princeton University
Press, Princeton, 1973), M. Gell-Mann (unpublished),
and the lucid discussion in R. Geroch,  No\^us,
18, 617, 1984.

\item {9.}  D. Finkelstein,   Trans. N.Y. Acad. Sci., 25, 621, 1963;
N. Graham, unpublished Ph.D. dissertation, University of North Carolina
1968 and in {\it The Many Worlds Interpretation of Quantum Mechanics}, ed.
by B. DeWitt, and N. Graham (Princeton University Press, Princeton, 1973);
and J. B. Hartle   Am. J. Phys., 36, 704, 1968.

\item {10.}  For a very clear discussion of this equivalence and its
consequences see C. M. Caves,   Phys. Rev., D33, 1643, 1986 and
ibid, D35, 1815, 1987. 

\item {11.}  H. Salecker and E. P. Wigner,  Phys. Rev., 109, 571, 1958.

\item {12.}  B. DeWitt,  Phys. Rev., 160, 1113, 1967.

\item {13.}  A. Peres,   Am. J. Phys., 48, 552, 1980.

\item {14.}  D. Page and W. Wooters,  Phys. Rev., D27, 2885, 1983, W.
 Wooters, 
Int. J. Th. Phys., 23, 701, 1984.

\item {15.}  See, e.g., J. B. Hartle and S. W. Hawking,  Phys. Rev.,
 D28, 2960, 1983,
S. W. Hawking (to be published), D. N. Page (to be published). J. B. Hartle,
Class. Quant. Grav., 2, 707, 1985, A. Anderson and B. DeWitt,
{\it Found. Phys.}, 16, 91, 1986. 

\item {16.}  See, e.g., H. Hamber and R. Williams,   Phys. Lett.,
157B, 368, 1985,   Nucl. Phys., B267, 482, 1986,   ibid,
B269, 712, 1986; H. Hamber in {\it Critical Phenomema, Random Systems and
Gauge Theories: Les Houches 1984} ed. by R. Stora and K. Osterwalder (Elsiever
Science Publishers, Amsterdam, 1986); J. B. Hartle,   J. Math. Phys.,
26, 804, 1985.

\item {17.}  H. Leutwyler,  Phys. Rev., 134, B1155, 1964, B. S. DeWitt,
 in {\it Magic
Without Magic:  John Archibald Wheeler}, ed. by J. Klauder, (Freeman, San
Francisco, 1972), L. Faddeev, and V. Popov,   Usp. Fiz. Nauk., 111,
427, 1973   Sov. Phys. -Usp., 16, 777, 1974, E. Fradkin and G.
Vilkovisky,  Phys. Rev., D8, 4241, 1973, M. Kaku,  Phys. Rev.,
 D15, 1019, 1977.

\item {18.}  E.g., S. W. Hawking in {\it Relativity Groups and Topology
II}  ed. by B. DeWitt and R. Stora (Elsevier, Amsterdam, 1984).

\item {19.}  G. Gibbons, S. W. Hawking and M. Perry,  Nucl. Phys.,
B138, 141, 1978, J. B. Hartle and K. Schleich, 
 {\it The Conformal Rotation in Linearized Gravity} 
in {\it Quantum Field Theory and Quantum Statistics}, ed. by  I. A. 
Batalin, C. J. Isham and G. A. Vilkovisky), Adam Hilger, Bristol, 67-87,
(1987), 
K. Schleich, {\it Conformal Rotation in Perturbative Gravity}, Phys. Rev D 36, 2342-2363 (1987). 

\item {20.}  J.A. Wheeler in {\it Problemi dei fondamenti della fisica,
Scoula internazionale di fisica ``Enrico Fermi''} Corso 52 ed. by G. Toraldo
di Francia
(North Holland,
Amsterdam, 1979).

\item {21.}  T. Banks,   Nucl. Phys., B249, 332, 1985.

\item {22.}  S. W. Hawking and J. Halliwell,  Phys. Rev., D31, 1777,
 1985.

\item {23.}  P. D'Eath and J. Halliwell, {\it  Fermions in Quantum Cosmology}, Phys. Rev. D, 35, 1100 (1987). 

\item {24.}  R. Brout, G. Horwitz and D. Weil, {\it On the Onset of Time and Temperature in Cosmology}, Phys. Lett B, 192, 318 (1987). 

%\item {25.}  J.B.~Hartle, {\it Quantum Kinematics of Spacetime III. General Relativity}, (unpublished). 

\item {26.}  S. Hojman, K. Kuchar, and C. Teitelboim,   Ann. Phys.
(N.Y.), 76, 97, 1976.

\item {27.}  See, e.g., N. D. Birrell and P. C. W. Davies {\it Quantum Fields
in Curved Space} (Cambridge University Press, Cambridge, 1982).

\item {28.}  A. Barvinsky and V. N. Ponomariov,  Phys. Lett., 167B,
289, 1986, A. Barvinsky,   Phys. Lett., 175B, 401, 1986.

\item {29.}  C. Misner, K. Thorne, and J. A. Wheeler, {\it Gravitation}
(W. H. Freeman, San Francisco, 1970).
\end{description}

\end{document}